\documentclass{article}
\usepackage{arxiv}
\usepackage[affil-it]{authblk}
\usepackage[table,xcdraw]{xcolor}

\usepackage[english]{babel}
\usepackage{blindtext}
\usepackage[utf8]{inputenc} 
\usepackage[T1]{fontenc}    
\usepackage{url}            
\usepackage{amsfonts}       
\usepackage{graphicx}
\usepackage{amsmath}
\usepackage{amssymb}
\usepackage{multirow}
\usepackage{tikz}
\usepackage{pgfplots}
\usepackage{lineno}
\usepackage{color}
\usepackage{adjustbox}
\usepackage{pdflscape}
\usepackage{subfig}
\usepackage{float}

\usepackage[font=small,labelfont=bf]{caption}

\usepackage{lscape}

\title{Describing the effect of influential spreaders on the different sectors of Indian Market: A complex networks perspective}

\author[1,*]{Anwesha Sengupta} 
\author[2]{Shashankaditya Upadhyay}
\author[3]{Indranil Mukherjee} 
\author[2]{Prasanta K. Panigrahi}

\affil[1]{Department Applied Statistics, Maulana Abul Kalam Azad University of Technology, Haringhata, 741249, West Bengal, India}
\affil[2]{Department of Physical Sciences, Indian Institute of Science Education and Research, Kolkata, Mohanpur, 741246, West Bengal, India}
\affil[3]{School of Management Sciences, Maulana Abul Kalam Azad University of Technology, Haringhata, 741249, West Bengal, India}
\affil[*]{Corresponding author: anweshasengupta.math05@gmail.com}

\date{}
\newtheorem{thm}{Definition}
\begin{document}

\maketitle
\begin{abstract}
Market competition has a role which is directly or indirectly associated with influential effects of individual sectors on other sectors of the economy. The present work studies the relative position of a product in the market through the identification of influential spreaders and its corresponding effect on the other sectors of the market using complex network analysis during the pre-, in-, and post-crisis induced lockdown periods using daily data of NSE from December, 2019 to June, 2021. The existing approaches using different centrality measures failed to distinguish between the positive and negative influences of the different sectors in the market which act as spreaders. To obviate this problem, this paper presents an effective measure called LIEST (Local Influential Effects for Specific Target) that can examine the positive and negative influences separately with respect to any crisis period. LIEST considers the combined impact of all possible nodes which are at most three steps away from the specific targets for the networks. The essence of non-linearity in the network dynamics without considering single node effect becomes visible particularly in the proposed network.
\end{abstract}


\section{Introduction:}
The idea of determining the relative position of a product in the market has been an active area of research for the development of emerging markets \cite{burns1986_relative_position}. The key factors \cite{kotlar_1996_key_factor} that have been recognised for establishment of this relative position are the influential and non-influential spreaders in the market, important examples being the addiction behavior among teenagers \cite{addiction-2021}, viral advertising of a newly promoted item \cite{Viral-market-2002} etc. Motivated by these examples the present work attempts to rank such spreaders in the context of their effect on the Indian financial markets during the pre-, in-, and post-COVID 19 induced lockdown periods.\\

Several methodologies have evolved and been used over the years to analyze the behaviour of financial markets \cite{BSEdetrend_2018_IM, kirkpatrick-diff-app-study,silva_2016_finance_structure}. Complex networks, being one of them, has been used as an effective tool. Complex Networks are generic representations of complex system in which the underlying topology is a graph (a mathematical object) \cite{boccaletti_2006_complex_network,2012networks, 2013networks}.  Ever since the 2008 subprime crisis, many physicists and mathematicians have used these in their work, where typically the nodes represent the stocks or financial assets while the edges represent the possible interactions between them\cite{PB-RMT,majapa2016topology_financial_crisis_2008, GC}. Such interactions can be analysed by employing techniques such as correlation \cite{entropy2020perspective,Thai,tumminello_2007_correlation}, minimal spanning tree \cite{mantegna-MST}, transfer entropy \cite{darbellay-2000}, recurrence plots \cite{recurrence-2012}, non-linear dependencies \cite{Alexander-2017},  causality \cite{PCM-2020, CMS-2014,GC, causal} etc. These representations help to strengthen our understanding about the structural behavior of the market. In case of real world scenarios involving financial markets and networks between assets, positive (friendly) and negative (foe-like) relations do not ideally exist as demarcated but rather as an admixture of the two. This necessitates the identification of the target specific influential nodes in the network which might cause the development or decline of a product/initiative/organization. A typical example in the financial sector would be an insurance company which while promoting a new product in the market suddenly finds the competitor creating a negative impact either by adding by some additional discounts on their existing offers in payment of premiums, or through more inclusion in the coverage under the policy or through some additional promotions. Keeping this idea in mind, the aim of this study is to determine the influential spreaders for positive (where the users positively influence others in the network) as well as negative (where the users act as foes or competitors) networks derived from the original network.  In continuation, the target specific influential nodes in the network needs to be identified, which might be the causes behind the process of development or regression of such companies.\\

Researchers have studied financial markets using analytical tools. In 1999, Mantegna \& Stanley \cite{Mantegna_book}, used tools of statistical physics such as power laws, random walks and scaling behaviour for the analysis of financial markets.  L. Laloux et al.\cite{laloux-1999} emphasized on understanding the structural pattern of empirical correlation matrices using random matrix theory (RMT). Later on, new perspectives emerged such as those involving agent-based market models \cite{giardina-2003} and networks studies \cite{Newman_2003_Networks}. Correlation based networks has been applied as a tool to study the behavior of complex networks occurring in various disciplines including financial markets \cite{Structural,corr-bio,Spectral-Pan, Thai, return, NSE-sinha}.
Using complex networks and their underlying structural dynamics, financial markets have been extensively studied and analysed over a considerable period of time from both theoretical and practical angles. Sinha et al.\cite{NSE-sinha} proposed that the emerging market NSE (National Stock Exchange) and developed market NYSE (New York Stock Exchange) were similar on the basis of eigenvalue distribution of cross-correlation matrices though the emerging markets tended to be more correlated than the developed one.  Haluszczynski \cite{Alexander-2017} affirmed that the concept of centrality might be used as some kind of early warning by analyzing the strength of non-linear dependencies derived from surrogate data.  Pharasi \cite{Pharasi-2018} proposed an idea to identify the early indicators for the critical states of financial market crashes based on the cross correlation structural patterns. Kumari et al. \cite{ML-Stock} proposed some machine learning models e.g. Long-Short Term Memory(LSTM), Multi-Layer Perceptron (MLP) etc. which are capable of bringing about improvement in the predictive analysis of stock markets.\\

It must be mentioned here that the localized target spreading problem  \cite{LIM} and the analysis based on traditional centrality measures like Degree centrality \cite{stability-2015}, Eigen centrality \cite{Network-centrality-2019, stability-2015}, Betweenness centrality \cite{RWBC, Brain-connect-2010}, Harmonic centrality \cite{Brain-connect-2010}, K-shell decomposition  \cite{K-shell} may not be sufficient to fulfill the requirements for identifying the positive and negative influential effects on the specific targets in the network \cite{CC,qu_2022_identification, LIM}. While it is true that the vertices ranked higher as per different centrality indices have a higher degree of role to play in propagation/regulation of the information flow (in our case financial information) across the network on a global scale, these centrality measures may fail to capture the local influences posed by a vertex (companies in our case) on information flow across the network, relative or specific to a particular subset of vertices in the network. Thus one of the main objectives of this work is to overcome this challenge of finding vertex (company) specific influences drawn on or posed by a specific subset of vertices in the network.

The Brain Connectivity Toolbox (BCT) is readily available in the public domain \cite{Brain-connect-2010} which may also be used to study the structural dynamics through degree centralities of networks. Recently in 2021, Luan Y et al. \cite{NSP} proposed and used a new measure, Improved Centrality Index (ICC), based on closeness centrality and a semi-local iterative algorithm to analyze the effect of the number of shortest paths on the identification of the influential spreaders. This is motivated by the fact that even if the shortest path length between two nodes is quite large, the interaction effect can be significant if there is a large number of such shortest path lengths between these nodes. Guilbeault \cite{CC} constructed measures of complex path length and complex centrality to enhance the capability for identifying network structures and central individuals which are most suitable in spreading complex contagions. Some other instances in recent literature where the competition in the financial market has been studied can be found in \cite{burns1986_relative_position, korean_2020_investor}.  Upadhyay et al.\cite{GC} found that the causal network structure of the stock prices of distinguished companies in the New York Stock Exchange (NYSE) showed different patterns before, during, and after the 2007-08 global financial crisis. It was also shown that the market was driven by consumer staples and health-care sectors during the period of the crisis, while the recovery phase was dominated by Material, Healthcare, and Consumer discretionary sectors.\\

It is often (wrongly) interpreted that the most central nodes are always likely to be the most influential nodes in a network  \cite{K-shell}. Rather, the question arises that how the most influential subsets of nodes of the network can be identified, such that they influence the target nodes with a positive and negative impact in a distinct manner. Song et al. \cite{LIM} proposed the Local Influence Matrix (LIM) method to find the set of seed nodes with maximum positive influence on the localized targets while keeping the influence on the non-target nodes as little as possible in case of signed social networks. The method mentioned therein, while suitable for capturing the influential effects on the localized target over the entire network, does not take into account the positive and negative influences separately. As market competition has a role which is directly or indirectly associated with positive or negative influential effects over the network \cite{oecd2009competition}, it is necessary to consider the impacts of these two effects separately. In the present study, attempts have been made to separately consider the positive and negative influential effects of the nodes where we try to draw attention over the target specific set of nodes in line with the proposed LIM method. The importance of the present work stems from the fact that a novel method named (by the present authors) "Local Influential Effects for Specific Target" (LIEST) has been applied to address the above mentioned problem. This technique considers the combined impact of all possible nodes which are at most three steps away from the specific targets for the networks and separately structures the positive as well as the negative networks. The essence of nonlinearity in the network dynamics, without considering the single node effect, becomes particularly visible in the proposed method. The COVID 19 pandemic and the global lockdown that resulted from it created a health
catastrophe of unprecedented magnitude and also severely disrupted the socio-economic
fabric of the world. All areas of the economy suffered in its aftermath. The various sectors of
the Indian stock market also came under stress. Due to the inter-connectedness of the
different sectors, the effect of the shock spread across the capital market as a whole.
Although it is quite difficult to make a comprehensive assessment of this impact at this
juncture, it is still an instructive idea to use a complex networks approach to analyze the
spread of this sudden impact. It is also necessary to see the response of different sectors of the
market when the COVID-19 restrictions were being gradually lifted. The present study
analyzed the NSE daily datasets during pre-, in- and post-COVID 19 periods (December, 2019 - June, 2021) of the Healthcare, Financial, Consumer Defensive (CD), Automobile, Basic Materials, Technology, Industrial and Consumer Cyclical (CC) sectors and observed that in the very beginning of the Covid period, Financial, Healthcare and CD sectors positively influenced the entire market whereas Automobile, Industrial, Consumer Cyclical and Basic Materials sectors had negative influential effects over the market. It is expected that, in case of a company belonging to a particular sector, that experiences a positive influential effect over the market, would grow and sustain itself, while any negative influence would stymie the growth process of an organization\cite{oecd2009competition}. The comparison of the present method (LIEST) with those employing traditional centrality measures, reveals its greater efficacy in real-life scenarios.

\subsection{General Overview:}
Section \ref{sec: Methods} describes the basic terminology, describes the datasets used  and provides the theory underlying the technique named as Local Influential Effects for Specific Target (LIEST). This includes few basic procedures pertaining to the global centrality indices, brief description about the daily closing prices of different stocks in NSE of India and the methodology of the proposed model along with its flow-chart. Section \ref{sec: result} presents the results obtained by applying the method to the daily NSE datasets. Section \ref{sec: conclude} provides a summary of the results obtained, draws certain conclusions and indicates the directions in which future studies may be undertake in this area.\\

\section{Material and Methods:}
\label{sec: Methods}

\subsection{Description:}
Let $G(V,E,W)$ be a graph, $V$ denotes the set of vertices and each edge ($e_{uv}$) is associated with an weight ($W_{uv}$), representing the strength of the interaction between the pair of vertices. Centrality is an important property of complex networks that determines the behaviour of dynamic processes. There are several metrics to quantify the node centrality in complex networks. Some of these measures are briefly discussed below.

\begin{thm}
    \textbf{Geodesics path:}
    A path, in graph theory, is defined as a sequence of finite or infinite number of edges, which joins the vertices without any repetition. A path containing a minimum number of edges is known as geodesic path or shortest path. For weighted graphs the minimum number of edges is replaced by minimum sum of edge weights.
\end{thm}

\begin{thm}
    \textbf{Degree Centrality \cite{stability-2015}:}
    It is the most simple centrality measure, which is defined as the number of links incident upon a node (i.e the number of ties that a node has). For a directed network, two separate measures of degree centrality are defined, viz. indegree and outdegree. For an unweighted graph the degree centrality of a vertex simply counts the number of edges connected to it, whereas for weighted graphs it calculates the sum of its weighted edges. Thus for a weighted graph, the degree centrality is defined by the sum of edge weights incident to each vertex. Mathematically, $$DC(v) = \sum_{u| (u,v) \in E}W_{uv}$$

\end{thm}

\begin{thm}

 \textbf{Betweenness Centrality (BC) \cite{Network-centrality-2019}:}
    Betweenness centrality attempts to detect the amount of influence a node has over the flow of information in a graph. In a network in which the flow is entirely or at least mostly along geodesic paths, the betweenness of a vertex measures how much flow will pass through that particular vertex. Betweenness centrality ranks the globally influential nodes in the networks by counting the number of shortest paths that go through that particular vertex.  In the present study, we use the betweenness centrality measure for weighted adjacency matrix by,

    $$BC(v)= \sum_{s \neq t}\frac{\sigma_{st}(v)}{\sigma_{st}}$$ where $\sigma_{st}(v)$ denotes the number of shortest path between $s$ and $t$ passes through $v$ and $\sigma_{st}$ denotes total number of shortest path between $s$ and $t$.
     Here, the number of shortest path has been considered by replacing the weights ($W_{uv}$) as  $W'_{uv} = 1-W_{uv}$, as the matrix elements $(W_{uv})$ denote the similarity value and lie within 0 and 1 [see. Section \ref{sec: Methodology}].
\end{thm}

\begin{thm}
   \textbf{Harmonic Centrality (HC) \cite{rochat2009harmonic, stability-2015}:}
   Harmonic Centrality is a variant of Closeness Centrality that was invented to solve the problem, the original formula dealt with unconnected graphs. \\
   HC is defined as the sum of the inverted distances instead of the inverted sum of the distances.
    \begin{equation} \label{eq}
    \begin{split}
    HC(v_i) &= \sum_{j \neq i} {\frac{1}{dist(v_i,v_j)} } 
    \end{split}
    \end{equation}
    Thus it avoids cases where an infinite dist outweighs the others.\\
    HC can be normalized by dividing by $(N-1)$, where $N$ be the number of nodes in the graph.
    \begin{equation} \label{eq}
    \begin{split}
    HC(v_i)_{normal} &= \frac{1}{N-1}\sum_{j \neq i} {\frac{1}{dist(v_i,v_j)} } 
    \end{split}
    \end{equation}

    Here, the matrix elements of $S_P$ and $S_N$ denote the similarity value lie within 0 and 1 [see. Section \ref{sec: Methodology}]. Harmonic centrality considers the length of the shortest path between the pair of vertices. So, here we use the weight values ($W_{v_i v_j}'$) as dissimilarity values [i.e., $W'_{v_i v_j} = 1-W_{v_i v_j}$] to capture the shortest path.

\end{thm}

\begin{thm}
    {\textbf{Random-Walk Harmonic Centrality (RWHC) \cite{CTD, RWHC}:}}
    Random walk harmonic centrality is a measure that describes how fast the information flows from a vertex to others in the network. The measure is captured by the expected length of a random walk\cite{ross1996stochastic} instead of taking the length of the shortest path \cite{RWHC}.

    \begin{equation} \label{eq}
    \begin{split}
    RWHC(v) &=\frac{1}{N}. \frac{1}{\sum_{u=1}^N H_{uv}}, if \hspace{2mm}v \hspace{2mm} is \hspace{2mm} not \hspace{2mm} an \hspace{2mm} isolated \hspace{2mm} vertex\\
     &= 0, otherwise
    \end{split}
    \end{equation}
    where, $H_{uv}$ be the mean hitting time \cite{CTD} between vertices $u$ and $v$.
\end{thm}

\begin{thm}
     {\textbf{Random-Walk Betweenness Centrality (RWBC)\cite{RWBC}:}}
    RWBC is a centrality measure that captures the influence of a node over the information spread throughout the network without considering the conventional approaches. This measure is suitable for a network in which information essentially moves about at random until it locates its target. It contains contributions from many paths that are not optimal, though shorter paths still dominate over the longer ones since a random walk is supposed to determine its target before long.

    This measure counts the number of times a node is traversed by a random walk \cite{ross1996stochastic} between two other nodes. Instead of considering only the geodesics paths, it takes into account the contribution of all possible random paths between the pair of nodes, though it still gives more weights to the geodesics path.

    RWBC has been calculated as expected net number of times a random walk passes through vertex $w$ on its way from a source $u$ to a target $v$ and has been averaged over all $u$ and $v$, and is considered for each separate component of the graph of interest.
\end{thm}

\begin{thm}
    {\textbf{Weighted Eigenvector Centrality \cite{Network-centrality-2019,stability-2015}:}}
    The eigenvector centrality EC, is an extension of degree centrality. It does not only consider the number of neighbors connected to it, but rather takes into account the neighbors of its neighbors. Thus the importance of a node is determined as a function of the importance of its neighbours. The importance of its neighbours, in turn, depends on how important their neighbours are, and so on. It may well be possible that a node with a few important neighbours has larger eigenvalue centrality than a node with different neighbours of limited importance. Let $G(V,E,W)$ be a graph, $V$ denotes the stocks, and each edge ($e_{uv}$) is associated with a weight ($W_{uv}$) representing the similarity value. Let $A$ be the adjacency matrix associated with the similarity values. The eigenvector centrality is defined by,\\

    \begin{equation} \label{eq EC}
        \begin{split}
            EC_u = \frac{1}{\lambda_{max}} \sum_{v \in N(u)} A_{uv}. EC_v
        \end{split}
    \end{equation}
     In (\ref{eq EC}), the weighted centrality value of a node $u$ is considered as a weighted average of the centrality values of its neighbors, where $ EC = \left[ EC_{v_1}, EC_{v_2}, \dots, EC_{v_n} \right]^T$, denotes the eigenvector corresponding to largest eigenvalue $\lambda_{max}$ of adjacency matrix $A$.
\end{thm}

\begin{thm}
    \textbf{Jaccard Index (JI) \cite{jaccard_1912}:}
    The Jaccard Index compares members of two sets to determine which members are shared and which are distinct. Thus it measures the similarity between two finite sets and is defined as the ratio of the cardinality of the intersection of these two to the cardinality of their union. Mathematically, it can be expressed as,
    \begin{equation} \label{eq}
        \begin{split}
           J(S,T) = \frac{|S \cap T|}{|S \cup T|}
        \end{split}
    \end{equation}
    where, $0 \leq J(S,T) \leq 1$.

The Jaccard distance is a measure of how dissimilar two sets are. It is the complement of the Jaccard index and is obtained by subtracting the Jaccard index from 1. \end{thm}

\subsection{Data set:}
  In this study the daily closing prices of 175 frequently traded stocks in the National Stock Exchange (NSE) of India for the period December, 2019 - June, 2021, available in the websites \url{http://nseindia.com/} and \url{https://finance.yahoo.com/} have been used. A trade-off between long and short time intervals has been considered while analysing the entire time period under review. It is common knowledge that too long a time interval may smoothen out most of the fluctuations leading to a great deal of $"coarse-graining"$ effect. On the other hand, a very short window may either not indicate the desired changes or capture the occasional large spikes only. Thus achieving an optimality is an important challenge for the researcher. In this analysis we have considered the ratio $\lambda \approx 2N$ [$N =$ number of stocks, $\lambda =$ length of each time period], and consider mainly, $N < \lambda < 3N$, which has been used extensively in the finance literature to achieve the trade-off mentioned previously\cite{Structural}.\\

 We have used 209 trading days from December 2019 to August 2020, namely the pre- and In-Covid periods while the other sub period includes 225 trading days from September 2020 to June 2021, which may be denoted as the Post-Covid period.

\subsection{Methodology:}
\label{sec: Methodology}
The logarithmic return of stock prices is found to follow a normal distribution \cite{return}. Hence, the cross-correlation matrices from log-return of 175 NSE stocks for two different time spans $(Pre- \hspace{1mm} and \hspace{1mm}In-Covid\hspace{1mm} and\hspace{1mm} Post-Covid\hspace{1mm} times)$ have been constructed. The daily log-returns for stock $i$ is defined by $$R'_i(t) = ln \left[\frac{P'_i(t)}{P'_i(t-1)} \right]$$
where $P'_i(t)$ denotes the daily closing price for the stock $i$ at time $t$.

Let $G(V,E,W)$ be a graph,the vertices $V$ denote the stocks, $|V|=175$ and each edge ($e_{ij}$) is associated with a weight ($W_{ij}$), representing the correlation value $(c_{ij})$ between the pair of stocks. The correlation network, constructed from the cross-correlation matrix, is a signed, weighted and undirected network. \\

We use the Markov chain concept for the random occurrences, that cause transitions from one state to another. Let, $x(t)$ denote the random variable that contains the state of the Markov chain at time $t$. Therefore, $x(t)= u_i$ implies the Markov chain is in state $u_i$. In case of a graph model, the state-space is the collection of all vertices that are contained in the vertex set $V$. The random walker will have the following restriction; it can make a transition only to an adjacent vertex with some transition probability \cite{CTD}. If $x(t)=u_i,$ the random walker can make a transition to an adjacent vertex $u_j$ with probability $p_{ij}= P(x(t+1)=u_j | x(t)=u_i)=\frac{|c_{ij}|}{d_i} $, where, $d_i = \sum_{j=1}^N|c_{ij}|$. The influence of node $i$ over node $j$ is given by $r_{ij} \in \left\{ -1,0,1\right\}$, and it is determined by the sign of correlation coefficients. Hence, the weighted adjacency matrix is given by $A = [A_{ij}]$, where $A_{ij} = p_{ij}.r_{ij}$.\\

The basic idea is to compute all positive \& negative paths (i.e., all the combined correlation effects of nodes which are at most 3 steps away from the original node) from $V-T$ to target node $T$ using $k=3$ intermediate steps \cite{LIM}, that denotes the probability of a node in $V-T$ which activates the node in $T$ and that can also be used for finding the influential nodes. Mathematically, $A_{ij}^k $ denotes the probability that node $i$ reaches node $j$ in $k$ steps \cite{ross1996stochastic}. So, it measures the co-movement of $i$ and $j^{th}$ nodes along with the effect of other adjacent vertices which are $k$ steps away. The elements of the main diagonal of the matrix $A^k$ represent a cycle, so we should exclude them to remove the self-influential cases. Let, $\Bar{A}^k$ be the new matrix, whose main diagonal elements are all 0. Now,
\begin{align}
    A^{k+1}=A \times \Bar{A}^k
\end{align}
Then the matrix having the combined effect using at most $k$ steps away between any pair of nodes can be calculated as,
\begin{align}
    S = \sum_{l=0}^k\Bar{A}^{l+1}
\end{align}

Now, let us split the matrix $S$ into two sub matrices, $S_P$ and $S_N$ as given below,\\

\begin{equation} \label{eq}
\begin{split}
S_P(i,j) &= S(i,j), if \hspace{1mm}S(i,j) >0\\
     &= 0, otherwise
\end{split}
\end{equation}

\begin{equation} \label{eq}
\begin{split}
     S_N(i,j) &= -S(i,j), if \hspace{1mm}S(i,j) <0\\
     &= 0, otherwise
\end{split}
\end{equation}

$S_P(i,j)$ (or, $S_N(i,j)$) measures the possibility that node $i$ activates node $j$ positively (or, negatively)\cite{LIM}.
The networks constructed from $S_P$ and $S_N$ are known as positive and negative networks respectively. In network $S_P$  $(S_N)$, the edges represent the positive (negative) influences between the pair of nodes. Our aim is to identify the most influential nodes in the positive network which can positively activate the given  targets, while activating the non-target nodes to non-target nodes as well as the intra-specific competition impact between the target nodes as little as possible. Similarly, for negative network, while identifying the most influential nodes that negatively affect the target nodes, we must try to reduce the  impact of non-target nodes to non target nodes as much as possible and increase the impact of intra-specific competition between the target nodes.\\

In the present study, the target nodes are specific to each different sector. The target node vector$(f)$ is given by a $|V| \times 1$ vector, whose particular sectoral indices are represented by 1 and all the other entries are 0. Similarly, the non target node vectors are denoted by $(f')$, where all the non target entries are 1 and all the target entries are 0. The positive (negative) influence of non-target nodes to target nodes is represented by $S_P.f$  $(S_N.f)$, whereas the positive (negative) influence of non-target nodes to non-target nodes is represented by $S_P.f'$ $(S_N.f')$. The intra-specific competition matrix is given by,

\begin{equation} \label{eq}
\begin{split}
S_P'(i,j) &= S_P(i,j), if \hspace{1mm}i \hspace{1mm} or \hspace{1mm}j \in T\\
    &= 0, otherwise
\end{split}
\end{equation}

\begin{equation} \label{eq}
\begin{split}
    S_N'(i,j) &= S_N(i,j), if \hspace{1mm}i \hspace{1mm} or \hspace{1mm}j \in T\\
    &= 0, otherwise
\end{split}
\end{equation}

Based on the above notations, we propose the following relational formula to capture the influential impacts on specific target nodes for both the networks which is given by, \\
\begin{align}
    S_{LIEST}^+ =  S_P.f -\beta_0.S_P.f' - \gamma_0.S_P'.f
\end{align}

\begin{align}
    S_{LIEST}^- =  S_N.f -\beta_1.S_N.f' + \gamma_1.S_N'.f
\end{align}
Based on the influential impacts, we can find the most significant nodes for each specific target sector. $S_{LIEST}^+$ denotes the positive influential effects for the specific set of target nodes in order of preferences. Mathematically, the entire effect measured from $S_{LIEST}^+$,  mostly depends on the effect of non-target nodes towards target nodes, while minimizing the effect of non-target nodes towards the non-target nodes and the intra-specific competition between the target nodes. On-the-other-hand, $S_{LIEST}^-$ denotes the negative influential effects for those specific set of target nodes, which highly depends on the effect of non-target nodes towards the target nodes and the intra-specific competition between the same, while minimizing the effect of non-target nodes on the non-target nodes. The effect of non-target nodes on the non-target nodes and intra-specific competition cannot be counted to their full extent regarding the contribution to $S_{LIEST}^+$ \& $S_{LIEST}^-$ and the amount of their reduced effects can be expressed when multiplied by the factors $\beta_0, \gamma_0, \beta_1, \gamma_1$ such that, $0 < \beta_0, \gamma_0, \beta_1, \gamma_1 < 1.$ The hyperparameters $\beta_0, \gamma_0, \beta_1, \gamma_1$ are chosen by observing the Jaccard Index (JI) and the positive values of influential effects in all cases to ensure their stability as well. The JI for two points have been obtained by comparing the  two neighboring points  in the $\beta-\gamma$ plane.

\begin{center}
 \includegraphics[width = 0.7 \textwidth]{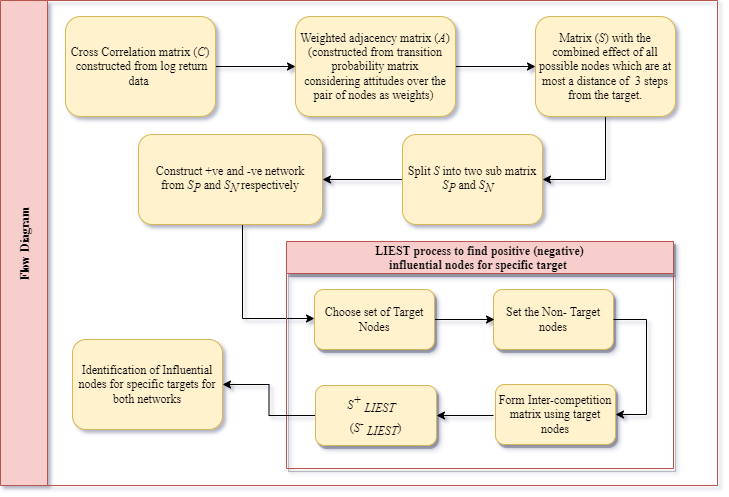}
 \captionof{figure}{Methodology}
\end{center}

\section{Result Analysis:}
\label{sec: result}
The derived networks resulting from the original correlation network capture the inherent nonlinearity of the data.
The LIEST process has been illustrated for the two separate networks and corresponding results have been tabulated (see table \ref{table 1}- \ref{table 4}). Here, Jaccard Similarity index and the positive values of influential effect for each case have been considered to specify the hyper parameters for each period in the different networks. For positive network, $\beta_0$ \& $\gamma_0$ have been set as 0.02 and 0.1 respectively while for negative network $\beta_1$ \& $\gamma_1$ have been considered to be 0.1 and 0.7 respectively.\\

The results obtained are found to be in agreement with real world observations\cite{GC}. In the beginning of the Covid period, the Healthcare $(HC)$ sector was positively self influential while there was also influence of Consumer Defensive $(CD)$ sectors. In the post-COVID period, the positive influence of Technology sectors has been significant along with the influence of Healthcare and $CD$ sectors. The positive influence of Automobile, Industrial and Consumer Cyclical$(CC)$ sectors was observed to have significant effects on the Financial sector during the in-COVID period as there was a shortage of cash flow in that period. On the contrary, the effects have been different in the post-COVID period. The self influential effects along with positive influence of $CD$ sectors and partial effects of Technology, Industrial and $CC$ sectors on the Financial sector have been noticed during the post-COVID recovery phase. In the in-COVID period, the $CD$ sector was positively influenced by mostly the Healthcare sector and partially by $CD$ and Energy sectors. The situation changed in the post-COVID era and the Consumer Defensive sector was highly influenced by Consumer Defensive, Healthcare, Basic materials $(BM)$ as well as Financial sectors. In the in-COVID period, it has been noticed that the Technology sector has been positively influenced by Automobile, Industrial, $CD$ and Automobile sectors whereas in the post-COVID scenario it has been positively influenced by mostly Healthcare, Technology, Financial and $CD$ sectors. The Automobile sector in the In-COVID time seemed to be positively influenced by the Financial, Automobile and Technology sectors whereas in the post-COVID era, Finance, Consumer Defensive and Industrial sectors have influenced the Automobile sector. During the in-COVID situation,it has been seen that the Basic Materials sector was positively influenced by Consumer Defensive, Industrial, Automobile and Financial sectors, but in the post-COVID period the Utilities, Industrial, Consumer Defensive and Consumer Cyclical sectors have positively influenced the same. The Industrial sector has been positively influenced by the Automobile sector, Financial and Consumer Cyclical sectors during in-COVID period. In post-COVID time the significance of Automobile sector has been slightly reduced and it is positively influenced by Automobile, Consumer Defensive, Finance \& Industrial sectors.\\

In the COVID period, the Automobile sector was observed to be negatively influenced by Healthcare, Consumer Defensive and Industrial sectors whereas, in the post-COVID period the Technology, Industrial, Consumer Cyclical and Basic materials sectors negatively influenced the Automobile sector along with its own negative self influential effect. During Covid time, the Technology sector was found to be negatively influenced by mostly Healthcare, Consumer Defensive, Industrial, Basic Material and Telecom sectors, but in Post-COVID time, the Healthcare, Automobile and Finance sectors exercised a negative influential effect on the Technology sector. It has been surprisingly observed that Automobile, Technology and Industrial sectors show negative influential impact on Basic Materials sector during the Covid period as well as in the post-COVID recovery phase. The negative influence of Automobile, Consumer Cyclical, Basic Materials have been observed along with negative self influential effect on the Industrial sector whereas in the Post-COVID time, Consumer Cyclical, Automobile, Technology, Healthcare and its own negative influential impact have been noticed on the Industrial sector. In the beginning of the Covid period, the negative influential effect of Industrial, Automobile, Healthcare and Consumer Defensive sectors have been observed on the Consumer Cyclical sector, whereas in the Post-COVID time Automobile, Consumer Cyclical, Healthcare and Finance sectors have together caused a negative effect  on the Consumer Cyclical sector.

\begin{landscape}
\begin{table}[h]
\caption{Positive influential nodes for different specific target sectors for positive network during in-COVID time, choosing $\beta_0 =0.02, \gamma_0= 0.1$}
\label{table 1}
\resizebox{\columnwidth}{!}{
\begin{tabular}{|l|llllllll|}
\hline
                                                             & \multicolumn{8}{c|}{\textbf{Influential nodes for Different specific target sectors for positive network for In covid time}}                                                                                                                                                                                                                                                                                                                                                                                                                              \\[1ex] \hline
                                                             & \multicolumn{1}{l|}{\textbf{Industrial (Ind)}}                    & \multicolumn{1}{l|}{\textbf{Financial}}                             & \multicolumn{1}{l|}{\textbf{Healthcare (HC)}}                      & \multicolumn{1}{l|}{\textbf{Automobile  (Auto)}}                     & \multicolumn{1}{l|}{\textbf{Consumer Defensive (CD)}}                & \multicolumn{1}{l|}{\textbf{Consumer Cyclical  (CC)}}               & \multicolumn{1}{l|}{\textbf{Basic Materials (BM)}}                   & \textbf{Technology}                             \\ \cline{2-9}
                                                             & \multicolumn{1}{l|}{\cellcolor[HTML]{FFFFFF}NAGAFERT - BM}        & \multicolumn{1}{l|}{\cellcolor[HTML]{FFFFFF}ARVIND - CC}            & \multicolumn{1}{l|}{\cellcolor[HTML]{FFFFFF}LUPIN.NS - HC}         & \multicolumn{1}{l|}{\cellcolor[HTML]{FFFFFF}SIYSIL - CC}             & \multicolumn{1}{l|}{\cellcolor[HTML]{FFFFFF}TORNTPHARM.NS - HC}      & \multicolumn{1}{l|}{\cellcolor[HTML]{FFFFFF}UNITECH - Real estate}  & \multicolumn{1}{l|}{\cellcolor[HTML]{FFFFFF}KAKATCEM - BM}           & \cellcolor[HTML]{FFFFFF}NESTLEIND.NS - CD       \\ \cline{2-9}
                                                             & \multicolumn{1}{l|}{\cellcolor[HTML]{FFFFFF}SUNDRMFAST - Auto}    & \multicolumn{1}{l|}{\cellcolor[HTML]{FFFFFF}EICHERMOT - Auto}       & \multicolumn{1}{l|}{\cellcolor[HTML]{FFFFFF}IPCALAB - HC}          & \multicolumn{1}{l|}{\cellcolor[HTML]{FFFFFF}BATAINDIA - CC}          & \multicolumn{1}{l|}{\cellcolor[HTML]{FFFFFF}DRREDDY - HC}            & \multicolumn{1}{l|}{\cellcolor[HTML]{FFFFFF}MAHSCOOTER - Auto}      & \multicolumn{1}{l|}{\cellcolor[HTML]{FFFFFF}BIRLACABLE - Technology} & \cellcolor[HTML]{FFFFFF}THERMAX - Industrial    \\ \cline{2-9}
                                                             & \multicolumn{1}{l|}{\cellcolor[HTML]{FFFFFF}BHARATFORG - Auto}    & \multicolumn{1}{l|}{\cellcolor[HTML]{FFFFFF}BIRLACABLE -Technology} & \multicolumn{1}{l|}{\cellcolor[HTML]{FFFFFF}CADILAHC.NS - HC}      & \multicolumn{1}{l|}{\cellcolor[HTML]{FFFFFF}ASHOKLEY- Auto}          & \multicolumn{1}{l|}{\cellcolor[HTML]{FFFFFF}IPCALAB - HC}            & \multicolumn{1}{l|}{\cellcolor[HTML]{FFFFFF}HINDMOTORS - Auto}      & \multicolumn{1}{l|}{\cellcolor[HTML]{FFFFFF}GIPCL - Utilities}       & \cellcolor[HTML]{FFFFFF}SUNDRMFAST - Auto       \\ \cline{2-9}
                                                             & \multicolumn{1}{l|}{\cellcolor[HTML]{FFFFFF}FMGOETZE - Auto}      & \multicolumn{1}{l|}{\cellcolor[HTML]{FFFFFF}BHARATFORG - Auto}      & \multicolumn{1}{l|}{\cellcolor[HTML]{FFFFFF}CIPLA - HC}            & \multicolumn{1}{l|}{\cellcolor[HTML]{FFFFFF}GNFC - BM}               & \multicolumn{1}{l|}{\cellcolor[HTML]{FFFFFF}CADILAHC.NS - HC}        & \multicolumn{1}{l|}{\cellcolor[HTML]{FFFFFF}ASHOKLEY - Auto}        & \multicolumn{1}{l|}{\cellcolor[HTML]{FFFFFF}HARRMALAYA - CD}         & \cellcolor[HTML]{FFFFFF}TATACOMM.NS  - Telecom  \\ \cline{2-9}
                                                             & \multicolumn{1}{l|}{\cellcolor[HTML]{FFFFFF}ROLTA - Technology}   & \multicolumn{1}{l|}{\cellcolor[HTML]{FFFFFF}MAHSEAMLES - CC}        & \multicolumn{1}{l|}{\cellcolor[HTML]{FFFFFF}TORNTPHARM.NS - HC}    & \multicolumn{1}{l|}{\cellcolor[HTML]{FFFFFF}MAHSCOOTER - Auto}       & \multicolumn{1}{l|}{\cellcolor[HTML]{FFFFFF}BIOCON.NS - HC}          & \multicolumn{1}{l|}{\cellcolor[HTML]{FFFFFF}NAGAFERT - BM}          & \multicolumn{1}{l|}{\cellcolor[HTML]{FFFFFF}LUMAXIND - Auto}         & \cellcolor[HTML]{FFFFFF}BRITANNIA - CD          \\ \cline{2-9}
                                                             & \multicolumn{1}{l|}{\cellcolor[HTML]{FFFFFF}HINDMOTORS - Auto}    & \multicolumn{1}{l|}{\cellcolor[HTML]{FFFFFF}UNITECH - Real estate}  & \multicolumn{1}{l|}{\cellcolor[HTML]{FFFFFF}DRREDDY - HC}          & \multicolumn{1}{l|}{\cellcolor[HTML]{FFFFFF}UCALFUEL - Auto}         & \multicolumn{1}{l|}{\cellcolor[HTML]{FFFFFF}ABBOTINDIA - HC}         & \multicolumn{1}{l|}{\cellcolor[HTML]{FFFFFF}IFCI - Financial}       & \multicolumn{1}{l|}{\cellcolor[HTML]{FFFFFF}LICHSGFIN - Finance}     & \cellcolor[HTML]{FFFFFF}BHARATFORG - Auto       \\ \cline{2-9}
                                                             & \multicolumn{1}{l|}{\cellcolor[HTML]{FFFFFF}SWARAJENG - Auto}     & \multicolumn{1}{l|}{\cellcolor[HTML]{FFFFFF}TATAMOTORS.NS - Auto}   & \multicolumn{1}{l|}{\cellcolor[HTML]{FFFFFF}DIVISLAB.NS - HC}      & \multicolumn{1}{l|}{\cellcolor[HTML]{FFFFFF}BERGEPAINT - BM}         & \multicolumn{1}{l|}{\cellcolor[HTML]{FFFFFF}AUROPHARMA - HC}         & \multicolumn{1}{l|}{\cellcolor[HTML]{FFFFFF}ELGIEQUIP - Industrial} & \multicolumn{1}{l|}{\cellcolor[HTML]{FFFFFF}ITC - CD}                & \cellcolor[HTML]{FFFFFF}SUPREMEIND - Industrial \\ \cline{2-9}
                                                             & \multicolumn{1}{l|}{\cellcolor[HTML]{FFFFFF}PIDILITIND - BM}      & \multicolumn{1}{l|}{\cellcolor[HTML]{FFFFFF}VOLTAS - Industrial}    & \multicolumn{1}{l|}{\cellcolor[HTML]{FFFFFF}COLPAL - CD}           & \multicolumn{1}{l|}{\cellcolor[HTML]{FFFFFF}HGS - Industrial}        & \multicolumn{1}{l|}{\cellcolor[HTML]{FFFFFF}CIPLA - HC}              & \multicolumn{1}{l|}{\cellcolor[HTML]{FFFFFF}DCW - BM}               & \multicolumn{1}{l|}{\cellcolor[HTML]{FFFFFF}VSTIND - CD}             & \cellcolor[HTML]{FFFFFF}HINDUNILVR - CD         \\ \cline{2-9}
                                                             & \multicolumn{1}{l|}{\cellcolor[HTML]{FFFFFF}IFCI - Finance}       & \multicolumn{1}{l|}{\cellcolor[HTML]{FFFFFF}KEC - Industrial}       & \multicolumn{1}{l|}{\cellcolor[HTML]{FFFFFF}SUNPHARMA - HC}        & \multicolumn{1}{l|}{\cellcolor[HTML]{FFFFFF}INFY - Technology}       & \multicolumn{1}{l|}{\cellcolor[HTML]{FFFFFF}WIPRO - Technology}      & \multicolumn{1}{l|}{\cellcolor[HTML]{FFFFFF}FINCABLES - Industrial} & \multicolumn{1}{l|}{\cellcolor[HTML]{FFFFFF}MTNL - Telecom}          & \cellcolor[HTML]{FFFFFF}BLUEDART - Industrial   \\ \cline{2-9}
                                                             & \multicolumn{1}{l|}{\cellcolor[HTML]{FFFFFF}INDHOTEL - CC}        & \multicolumn{1}{l|}{\cellcolor[HTML]{FFFFFF}BOSCHLTD.NS - Auto}     & \multicolumn{1}{l|}{\cellcolor[HTML]{FFFFFF}BIOCON.NS - HC}        & \multicolumn{1}{l|}{\cellcolor[HTML]{FFFFFF}BAJAJ-AUTO - Auto}       & \multicolumn{1}{l|}{\cellcolor[HTML]{FFFFFF}DIVISLAB.NS - HC}        & \multicolumn{1}{l|}{\cellcolor[HTML]{FFFFFF}TIRUMALCHM - BM}        & \multicolumn{1}{l|}{\cellcolor[HTML]{FFFFFF}KSB - Industrial}        & \cellcolor[HTML]{FFFFFF}MOTHERSUMI.NS - Auto    \\ \cline{2-9}
                                                             & \multicolumn{1}{l|}{\cellcolor[HTML]{FFFFFF}BOSCHLTD.NS - Auto}   & \multicolumn{1}{l|}{\cellcolor[HTML]{FFFFFF}ZEEL - CD}              & \multicolumn{1}{l|}{\cellcolor[HTML]{FFFFFF}MCDOWELL-N.NS - CD}    & \multicolumn{1}{l|}{\cellcolor[HTML]{FFFFFF}CHOLAFIN - Finance}      & \multicolumn{1}{l|}{\cellcolor[HTML]{FFFFFF}COLPAL - CD}             & \multicolumn{1}{l|}{\cellcolor[HTML]{FFFFFF}SUPPETRO - BM}          & \multicolumn{1}{l|}{\cellcolor[HTML]{FFFFFF}HIMATSEIDE - CC}         & \cellcolor[HTML]{FFFFFF}SWARAJENG - Auto        \\ \cline{2-9}
                                                             & \multicolumn{1}{l|}{\cellcolor[HTML]{FFFFFF}THOMASCOOK - CC}      & \multicolumn{1}{l|}{\cellcolor[HTML]{FFFFFF}TATAMETALI - BM}        & \multicolumn{1}{l|}{\cellcolor[HTML]{FFFFFF}HINDUNILVR - CD}       & \multicolumn{1}{l|}{\cellcolor[HTML]{FFFFFF}SUPPETRO - BM}           & \multicolumn{1}{l|}{\cellcolor[HTML]{FFFFFF}COSMOFILMS - CC}         & \multicolumn{1}{l|}{\cellcolor[HTML]{FFFFFF}RELCAPITAL - Finance}   & \multicolumn{1}{l|}{\cellcolor[HTML]{FFFFFF}ADANIENT - Energy}       & \cellcolor[HTML]{FFFFFF}HINDALCO - BM           \\ \cline{2-9}
                                                             & \multicolumn{1}{l|}{\cellcolor[HTML]{FFFFFF}MIRCELECTR - CC}      & \multicolumn{1}{l|}{\cellcolor[HTML]{FFFFFF}SUNDRMFAST - Auto}      & \multicolumn{1}{l|}{\cellcolor[HTML]{FFFFFF}TATACOMM.NS - Telecom} & \multicolumn{1}{l|}{\cellcolor[HTML]{FFFFFF}MASTEK - Technology}     & \multicolumn{1}{l|}{\cellcolor[HTML]{FFFFFF}RELIANCE - Energy}       & \multicolumn{1}{l|}{\cellcolor[HTML]{FFFFFF}SURYAROSNI - BM}        & \multicolumn{1}{l|}{\cellcolor[HTML]{FFFFFF}DHAMPURSUG - CD}         & \cellcolor[HTML]{FFFFFF}RELIANCE - Energy       \\ \cline{2-9}
                                                             & \multicolumn{1}{l|}{\cellcolor[HTML]{FFFFFF}CANFINHOME - Finance} & \multicolumn{1}{l|}{\cellcolor[HTML]{FFFFFF}TITAN - CC}             & \multicolumn{1}{l|}{\cellcolor[HTML]{FFFFFF}ABBOTINDIA - HC}       & \multicolumn{1}{l|}{\cellcolor[HTML]{FFFFFF}MPHASIS.NS - Technology} & \multicolumn{1}{l|}{\cellcolor[HTML]{FFFFFF}SUNPHARMA - HC}          & \multicolumn{1}{l|}{\cellcolor[HTML]{FFFFFF}SWARAJENG - Auto}       & \multicolumn{1}{l|}{\cellcolor[HTML]{FFFFFF}CENTENKA - Industrial}   & \cellcolor[HTML]{FFFFFF}NTPC.NS - Energy        \\ \cline{2-9}
\multirow{-16}{*}{\rotatebox{90}{Considering Atmost top 15 positive values}} & \multicolumn{1}{l|}{\cellcolor[HTML]{FFFFFF}JCHAC.NS - CC}        & \multicolumn{1}{l|}{\cellcolor[HTML]{FFFFFF}TATAPOWER - Utilities}  & \multicolumn{1}{l|}{\cellcolor[HTML]{FFFFFF}AUROPHARMA - HC}       & \multicolumn{1}{l|}{\cellcolor[HTML]{FFFFFF}BAJFINANCE - Finance}    & \multicolumn{1}{l|}{\cellcolor[HTML]{FFFFFF}BHARTIARTL.NS - Telecom} & \multicolumn{1}{l|}{\cellcolor[HTML]{FFFFFF}KARURVYSYA - Finance}   & \multicolumn{1}{l|}{\cellcolor[HTML]{FFFFFF}MOTHERSUMI.NS - Auto}    & \cellcolor[HTML]{FFFFFF}ONGC - Energy           \\ \hline
\end{tabular}}
\end{table}

\begin{table}[h]
\caption{Negative influential nodes for different specific target sectors for negative network during in-COVID time, choosing $\beta_0 =0.1, \gamma_0= 0.7$ }
\label{table 2}
\resizebox{\columnwidth}{!}{
\begin{tabular}{|l|llllllll|}
\hline
                                                             & \multicolumn{8}{c|}{\textbf{Influential nodes for Different specific target sectors for negative network for In covid time}}                                                                                                                                                                                                                                                                                                                                                                                                                     \\[1ex] \hline
                                                             & \multicolumn{1}{l|}{\textbf{Industrial (Ind)}}                     & \multicolumn{1}{l|}{\textbf{Financial}}                           & \multicolumn{1}{l|}{\textbf{Healthcare (HC)}}                      & \multicolumn{1}{l|}{\textbf{Automobile  (Auto)}}                  & \multicolumn{1}{l|}{\textbf{Consumer Defensive (CD)}}                & \multicolumn{1}{l|}{\textbf{Consumer Cyclical  (CC)}}              & \multicolumn{1}{l|}{\textbf{Basic Materials (BM)}}                 & \textbf{Technology}                           \\ \cline{2-9}
                                                             & \multicolumn{1}{l|}{\cellcolor[HTML]{FFFFFF}RUCHI - CD}            & \multicolumn{1}{l|}{\cellcolor[HTML]{FFFFFF}ESCORTS - Industrial} & \multicolumn{1}{l|}{\cellcolor[HTML]{FFFFFF}RUCHI - CD}            & \multicolumn{1}{l|}{\cellcolor[HTML]{FFFFFF}BASF - BM}            & \multicolumn{1}{l|}{\cellcolor[HTML]{FFFFFF}ESCORTS - Industrial}    & \multicolumn{1}{l|}{\cellcolor[HTML]{FFFFFF}ESCORTS - Industrial}  & \multicolumn{1}{l|}{\cellcolor[HTML]{FFFFFF}RUCHI - CD}            & \cellcolor[HTML]{FFFFFF}BASF - BM             \\ \cline{2-9}
                                                             & \multicolumn{1}{l|}{\cellcolor[HTML]{FFFFFF}ESCORTS - Industrial}  & \multicolumn{1}{l|}{\cellcolor[HTML]{FFFFFF}INFY - Technology}    & \multicolumn{1}{l|}{\cellcolor[HTML]{FFFFFF}M\&M - Auto}           & \multicolumn{1}{l|}{\cellcolor[HTML]{FFFFFF}LUPIN.NS - HC}        & \multicolumn{1}{l|}{\cellcolor[HTML]{FFFFFF}SHYAMTEL - Technology}   & \multicolumn{1}{l|}{\cellcolor[HTML]{FFFFFF}SHYAMTEL - Technology} & \multicolumn{1}{l|}{\cellcolor[HTML]{FFFFFF}ICIL - CC}             & \cellcolor[HTML]{FFFFFF}SAKHTISUG - CD        \\ \cline{2-9}
                                                             & \multicolumn{1}{l|}{\cellcolor[HTML]{FFFFFF}ICIL - CC}             & \multicolumn{1}{l|}{\cellcolor[HTML]{FFFFFF}ICIL - CC}            & \multicolumn{1}{l|}{\cellcolor[HTML]{FFFFFF}BAJAJ-AUTO - Auto}     & \multicolumn{1}{l|}{\cellcolor[HTML]{FFFFFF}RUCHI - CD}           & \multicolumn{1}{l|}{\cellcolor[HTML]{FFFFFF}M\&M - Auto}             & \multicolumn{1}{l|}{\cellcolor[HTML]{FFFFFF}HGS - Industrial}      & \multicolumn{1}{l|}{\cellcolor[HTML]{FFFFFF}ESCORTS - Industrial}  & \cellcolor[HTML]{FFFFFF}LUPIN.NS - HC         \\ \cline{2-9}
                                                             & \multicolumn{1}{l|}{\cellcolor[HTML]{FFFFFF}M\&M - Auto}           & \multicolumn{1}{l|}{\cellcolor[HTML]{FFFFFF}M\&M - Auto}          & \multicolumn{1}{l|}{\cellcolor[HTML]{FFFFFF}UCALFUEL - Auto}       & \multicolumn{1}{l|}{\cellcolor[HTML]{FFFFFF}TATACOMM.NS- Telecom} & \multicolumn{1}{l|}{\cellcolor[HTML]{FFFFFF}RUCHI - CD}              & \multicolumn{1}{l|}{\cellcolor[HTML]{FFFFFF}M\&M - Auto}           & \multicolumn{1}{l|}{\cellcolor[HTML]{FFFFFF}M\&M - Auto}           & \cellcolor[HTML]{FFFFFF}UNITECH - Real estate \\ \cline{2-9}
                                                             & \multicolumn{1}{l|}{\cellcolor[HTML]{FFFFFF}HGS - Industrial}      & \multicolumn{1}{l|}{\cellcolor[HTML]{FFFFFF}}                     & \multicolumn{1}{l|}{\cellcolor[HTML]{FFFFFF}ICIL - CC}             & \multicolumn{1}{l|}{\cellcolor[HTML]{FFFFFF}BEML - Industrial}    & \multicolumn{1}{l|}{\cellcolor[HTML]{FFFFFF}BAJAJ-AUTO}              & \multicolumn{1}{l|}{\cellcolor[HTML]{FFFFFF}BASF - BM}             & \multicolumn{1}{l|}{\cellcolor[HTML]{FFFFFF}SHYAMTEL - Technology} & \cellcolor[HTML]{FFFFFF}ROLTA - Technology    \\ \cline{2-9}
                                                             & \multicolumn{1}{l|}{\cellcolor[HTML]{FFFFFF}BASF - BM}             & \multicolumn{1}{l|}{\cellcolor[HTML]{FFFFFF}}                     & \multicolumn{1}{l|}{\cellcolor[HTML]{FFFFFF}SHYAMTEL - Technology} & \multicolumn{1}{l|}{\cellcolor[HTML]{FFFFFF}DRREDDY - HC}         & \multicolumn{1}{l|}{\cellcolor[HTML]{FFFFFF}HGS - Industrial}        & \multicolumn{1}{l|}{\cellcolor[HTML]{FFFFFF}BERGEPAINT - BM}       & \multicolumn{1}{l|}{\cellcolor[HTML]{FFFFFF}INFY - Technology}     & \cellcolor[HTML]{FFFFFF}TATACOMM.NS - Telecom \\ \cline{2-9}
                                                             & \multicolumn{1}{l|}{\cellcolor[HTML]{FFFFFF}INFY - Technology}     & \multicolumn{1}{l|}{\cellcolor[HTML]{FFFFFF}}                     & \multicolumn{1}{l|}{\cellcolor[HTML]{FFFFFF}INFY - Technology}     & \multicolumn{1}{l|}{\cellcolor[HTML]{FFFFFF}CIPLA - HC}           & \multicolumn{1}{l|}{\cellcolor[HTML]{FFFFFF}UNITECH - Real estate}   & \multicolumn{1}{l|}{\cellcolor[HTML]{FFFFFF}BAJAJ-AUTO - Auto}     & \multicolumn{1}{l|}{\cellcolor[HTML]{FFFFFF}HGS - Industrial}      & \cellcolor[HTML]{FFFFFF}NAHARINDUS - CC       \\ \cline{2-9}
                                                             & \multicolumn{1}{l|}{\cellcolor[HTML]{FFFFFF}BERGEPAINT - BM}       & \multicolumn{1}{l|}{\cellcolor[HTML]{FFFFFF}}                     & \multicolumn{1}{l|}{\cellcolor[HTML]{FFFFFF}ESCORTS - Industrial}  & \multicolumn{1}{l|}{\cellcolor[HTML]{FFFFFF}ASIANHOTNR - CC}      & \multicolumn{1}{l|}{\cellcolor[HTML]{FFFFFF}PFIZER - HC}             & \multicolumn{1}{l|}{\cellcolor[HTML]{FFFFFF}RUCHI - CD}            & \multicolumn{1}{l|}{\cellcolor[HTML]{FFFFFF}BERGEPAINT - BM}       & \cellcolor[HTML]{FFFFFF}INDIACEM - BM         \\ \cline{2-9}
                                                             & \multicolumn{1}{l|}{\cellcolor[HTML]{FFFFFF}BAJAJ-AUTO - Auto}     & \multicolumn{1}{l|}{\cellcolor[HTML]{FFFFFF}}                     & \multicolumn{1}{l|}{\cellcolor[HTML]{FFFFFF}MAHSCOOTER - Auto}     & \multicolumn{1}{l|}{\cellcolor[HTML]{FFFFFF}INDIACEM - BM}        & \multicolumn{1}{l|}{\cellcolor[HTML]{FFFFFF}DRREDDY - HC}            & \multicolumn{1}{l|}{\cellcolor[HTML]{FFFFFF}TATACOMM.NS - Telecom} & \multicolumn{1}{l|}{\cellcolor[HTML]{FFFFFF}BAJAJ-AUTO - Auto}     & \cellcolor[HTML]{FFFFFF}GLAXO - HC            \\ \cline{2-9}
                                                             & \multicolumn{1}{l|}{\cellcolor[HTML]{FFFFFF}TATACOMM.NS - Telecom} & \multicolumn{1}{l|}{\cellcolor[HTML]{FFFFFF}}                     & \multicolumn{1}{l|}{\cellcolor[HTML]{FFFFFF}ASHOKLEY - Auto}       & \multicolumn{1}{l|}{\cellcolor[HTML]{FFFFFF}HARRMALAYA - CD}      & \multicolumn{1}{l|}{\cellcolor[HTML]{FFFFFF}DEEPAKFERT - BM}         & \multicolumn{1}{l|}{\cellcolor[HTML]{FFFFFF}ICIL - CC}             & \multicolumn{1}{l|}{\cellcolor[HTML]{FFFFFF}UCALFUEL - Auto}       & \cellcolor[HTML]{FFFFFF}RAMANEWS - BM         \\ \cline{2-9}
                                                             & \multicolumn{1}{l|}{\cellcolor[HTML]{FFFFFF}BEML - Industrial}     & \multicolumn{1}{l|}{\cellcolor[HTML]{FFFFFF}}                     & \multicolumn{1}{l|}{\cellcolor[HTML]{FFFFFF}BERGEPAINT - BM}       & \multicolumn{1}{l|}{\cellcolor[HTML]{FFFFFF}GLAXO - HC}           & \multicolumn{1}{l|}{\cellcolor[HTML]{FFFFFF}INGERRAND - Industrial}  & \multicolumn{1}{l|}{\cellcolor[HTML]{FFFFFF}LUPIN.NS - HC}         & \multicolumn{1}{l|}{\cellcolor[HTML]{FFFFFF}GNFC - BM}                  & \cellcolor[HTML]{FFFFFF}BHEL - Industrial     \\ \cline{2-9}
                                                             & \multicolumn{1}{l|}{\cellcolor[HTML]{FFFFFF}MAHSCOOTER - Auto}     & \multicolumn{1}{l|}{\cellcolor[HTML]{FFFFFF}}                     & \multicolumn{1}{l|}{\cellcolor[HTML]{FFFFFF}GNFC - BM}             & \multicolumn{1}{l|}{\cellcolor[HTML]{FFFFFF}SUNPHARMA - HC}       & \multicolumn{1}{l|}{\cellcolor[HTML]{FFFFFF}GODFRYPHLP - CD}         & \multicolumn{1}{l|}{\cellcolor[HTML]{FFFFFF}BEML - Industrial}     & \multicolumn{1}{l|}{\cellcolor[HTML]{FFFFFF}ASHOKLEY - Auto}       & \cellcolor[HTML]{FFFFFF}RAJSREESUG - CD       \\ \cline{2-9}
                                                             & \multicolumn{1}{l|}{\cellcolor[HTML]{FFFFFF}ASIANHOTNR - CC}       & \multicolumn{1}{l|}{\cellcolor[HTML]{FFFFFF}}                     & \multicolumn{1}{l|}{\cellcolor[HTML]{FFFFFF}BATAINDIA - CC}        & \multicolumn{1}{l|}{\cellcolor[HTML]{FFFFFF}RELCAPITAL - Finance} & \multicolumn{1}{l|}{\cellcolor[HTML]{FFFFFF}NAGAFERT - BM}           & \multicolumn{1}{l|}{\cellcolor[HTML]{FFFFFF}GLAXO - HC}            & \multicolumn{1}{l|}{\cellcolor[HTML]{FFFFFF}BASF - BM}                  & \cellcolor[HTML]{FFFFFF}BEML - Industrial     \\ \cline{2-9}
                                                             & \multicolumn{1}{l|}{\cellcolor[HTML]{FFFFFF}INDIACEM - BM}         & \multicolumn{1}{l|}{\cellcolor[HTML]{FFFFFF}}                     & \multicolumn{1}{l|}{\cellcolor[HTML]{FFFFFF}HGS - Industrial}      & \multicolumn{1}{l|}{\cellcolor[HTML]{FFFFFF}GESHIP - Industrial}  & \multicolumn{1}{l|}{\cellcolor[HTML]{FFFFFF}SUPREMEIND - Industrial} & \multicolumn{1}{l|}{\cellcolor[HTML]{FFFFFF}BHEL - Industrial}     & \multicolumn{1}{l|}{\cellcolor[HTML]{FFFFFF}MAHSCOOTER- Auto}      & \cellcolor[HTML]{FFFFFF}FOSECOIND - BM        \\ \cline{2-9}
\multirow{-16}{*}{\rotatebox{90}{Considering Atmost top 15 positive values}} & \multicolumn{1}{l|}{\cellcolor[HTML]{FFFFFF}RELCAPITAL - Finance}  & \multicolumn{1}{l|}{\cellcolor[HTML]{FFFFFF}}                     & \multicolumn{1}{l|}{\cellcolor[HTML]{FFFFFF}}                      & \multicolumn{1}{l|}{\cellcolor[HTML]{FFFFFF}NAHARINDUS - CC}      & \multicolumn{1}{l|}{\cellcolor[HTML]{FFFFFF}ESABINDIA - Industrial}  & \multicolumn{1}{l|}{\cellcolor[HTML]{FFFFFF}CRISIL - Finance}      & \multicolumn{1}{l|}{\cellcolor[HTML]{FFFFFF}TATACOMM.NS - Telecom}           & \cellcolor[HTML]{FFFFFF}MTNL - Telecom        \\ \hline
\end{tabular}}
\end{table}
\end{landscape}

\begin{landscape}
\begin{table}[h]
\caption{Positive influential nodes for different specific target sectors for positive network in post-COVID time, choosing $\beta_0 =0.02, \gamma_0= 0.1$ }
\label{table 3}
\resizebox{\columnwidth}{!}{
\begin{tabular}{|
>{\columncolor[HTML]{FFFFFF}}c |
>{\columncolor[HTML]{FFFFFF}}l
>{\columncolor[HTML]{FFFFFF}}l
>{\columncolor[HTML]{FFFFFF}}l
>{\columncolor[HTML]{FFFFFF}}l
>{\columncolor[HTML]{FFFFFF}}l
>{\columncolor[HTML]{FFFFFF}}l
>{\columncolor[HTML]{FFFFFF}}l
>{\columncolor[HTML]{FFFFFF}}l |}
\hline
\multicolumn{1}{|l|}{\cellcolor[HTML]{FFFFFF}}                & \multicolumn{8}{c|}{\cellcolor[HTML]{FFFFFF}\textbf{Influential   nodes for Different specific target sectors for positive network for post   covid time}}                                                                                                                                                                                                                                                                                                                                                                                                                            \\[1ex] \hline
\cellcolor[HTML]{FFFFFF}                                                               & \multicolumn{1}{l|}{\cellcolor[HTML]{FFFFFF}\textbf{Industrial   (Ind)}} & \multicolumn{1}{l|}{\cellcolor[HTML]{FFFFFF}\textbf{Financial}}        & \multicolumn{1}{l|}{\cellcolor[HTML]{FFFFFF}\textbf{Healthcare   (HC)}} & \multicolumn{1}{l|}{\cellcolor[HTML]{FFFFFF}\textbf{Automobile   (Auto)}} & \multicolumn{1}{l|}{\cellcolor[HTML]{FFFFFF}\textbf{Consumer   Defensive (CD)}} & \multicolumn{1}{l|}{\cellcolor[HTML]{FFFFFF}\textbf{Consumer   Cyclical (CC)}} & \multicolumn{1}{l|}{\cellcolor[HTML]{FFFFFF}\textbf{Basic   Materials (BM)}} & \textbf{Technology}       \\ \cline{2-9}
\cellcolor[HTML]{FFFFFF}                                                               & \multicolumn{1}{l|}{\cellcolor[HTML]{FFFFFF}SUPREMIND   - Industrial}    & \multicolumn{1}{l|}{\cellcolor[HTML]{FFFFFF}HDFC   - Finance}          & \multicolumn{1}{l|}{\cellcolor[HTML]{FFFFFF}DRREDDY   - HC}             & \multicolumn{1}{l|}{\cellcolor[HTML]{FFFFFF}ASHOKLEY   - Auto}            & \multicolumn{1}{l|}{\cellcolor[HTML]{FFFFFF}HINDUNILVR   - CD}                  & \multicolumn{1}{l|}{\cellcolor[HTML]{FFFFFF}UNITECH   - Real estate}           & \multicolumn{1}{l|}{\cellcolor[HTML]{FFFFFF}COSMOFILMS   - CC}               & MPHASIS.NS   - Technology \\ \cline{2-9}
\cellcolor[HTML]{FFFFFF}                                                               & \multicolumn{1}{l|}{\cellcolor[HTML]{FFFFFF}ELGIEQUIP   - Industrial}    & \multicolumn{1}{l|}{\cellcolor[HTML]{FFFFFF}HDFCBANK   - Finance}      & \multicolumn{1}{l|}{\cellcolor[HTML]{FFFFFF}CIPLA   - HC}               & \multicolumn{1}{l|}{\cellcolor[HTML]{FFFFFF}HDFCBANK   - Finance}         & \multicolumn{1}{l|}{\cellcolor[HTML]{FFFFFF}NESTLEIND.NS   - CD}                & \multicolumn{1}{l|}{\cellcolor[HTML]{FFFFFF}THOMASCOOK   - CC}                 & \multicolumn{1}{l|}{\cellcolor[HTML]{FFFFFF}HEG   - Industrial}              & DRREDDY   - HC            \\ \cline{2-9}
\cellcolor[HTML]{FFFFFF}                                                               & \multicolumn{1}{l|}{\cellcolor[HTML]{FFFFFF}APOLLOTYRE   - Auto}         & \multicolumn{1}{l|}{\cellcolor[HTML]{FFFFFF}KOTAKBANK.NS   - Finance}  & \multicolumn{1}{l|}{\cellcolor[HTML]{FFFFFF}LUPIN   - HC}               & \multicolumn{1}{l|}{\cellcolor[HTML]{FFFFFF}RUCHI   - CD}                 & \multicolumn{1}{l|}{\cellcolor[HTML]{FFFFFF}ASIANPAINT   - BM}                  & \multicolumn{1}{l|}{\cellcolor[HTML]{FFFFFF}FINCABLES   - Industrial}          & \multicolumn{1}{l|}{\cellcolor[HTML]{FFFFFF}EIDPARRY   - CD}                 & CIPLA   - HC              \\ \cline{2-9}
\cellcolor[HTML]{FFFFFF}                                                               & \multicolumn{1}{l|}{\cellcolor[HTML]{FFFFFF}JCHAC.NS   - CC}             & \multicolumn{1}{l|}{\cellcolor[HTML]{FFFFFF}LT.NS   - CD}              & \multicolumn{1}{l|}{\cellcolor[HTML]{FFFFFF}TORNTPHARM.NS   - HC}       & \multicolumn{1}{l|}{\cellcolor[HTML]{FFFFFF}HDFC   - Finance}             & \multicolumn{1}{l|}{\cellcolor[HTML]{FFFFFF}BRITANNIA   - CD}                   & \multicolumn{1}{l|}{\cellcolor[HTML]{FFFFFF}MTNL   - Telecom}                  & \multicolumn{1}{l|}{\cellcolor[HTML]{FFFFFF}CESC   - Utilities}              & TCS   - Technology        \\ \cline{2-9}
\cellcolor[HTML]{FFFFFF}                                                               & \multicolumn{1}{l|}{\cellcolor[HTML]{FFFFFF}SUNDRMFAST   - Auto}         & \multicolumn{1}{l|}{\cellcolor[HTML]{FFFFFF}ITC   - CD}                & \multicolumn{1}{l|}{\cellcolor[HTML]{FFFFFF}CADILAHC.NS   - HC}         & \multicolumn{1}{l|}{\cellcolor[HTML]{FFFFFF}LT.NS   - CD}                 & \multicolumn{1}{l|}{\cellcolor[HTML]{FFFFFF}DABUR   - CD}                       & \multicolumn{1}{l|}{\cellcolor[HTML]{FFFFFF}MAHSCOOTER-   Auto}                & \multicolumn{1}{l|}{\cellcolor[HTML]{FFFFFF}VOLTAS   - Industrial}           & BERGERPAINT   - BM        \\ \cline{2-9}
\cellcolor[HTML]{FFFFFF}                                                               & \multicolumn{1}{l|}{\cellcolor[HTML]{FFFFFF}MARALOVER   - CC}            & \multicolumn{1}{l|}{\cellcolor[HTML]{FFFFFF}RAMANEWS   - BM}           & \multicolumn{1}{l|}{\cellcolor[HTML]{FFFFFF}AUROPHARMA   - HC}          & \multicolumn{1}{l|}{\cellcolor[HTML]{FFFFFF}KEC   - Industrial}           & \multicolumn{1}{l|}{\cellcolor[HTML]{FFFFFF}POLYPLEX-   BM}                     & \multicolumn{1}{l|}{\cellcolor[HTML]{FFFFFF}BEML   - Industrial}               & \multicolumn{1}{l|}{\cellcolor[HTML]{FFFFFF}HARRMALAYA   - CD}               & WIPRO   - Tecchnology     \\ \cline{2-9}
\cellcolor[HTML]{FFFFFF}                                                               & \multicolumn{1}{l|}{\cellcolor[HTML]{FFFFFF}NAGAFERT   - BM}             & \multicolumn{1}{l|}{\cellcolor[HTML]{FFFFFF}MTNL   - Telecom}          & \multicolumn{1}{l|}{\cellcolor[HTML]{FFFFFF}IPCALAB   - HC}             & \multicolumn{1}{l|}{\cellcolor[HTML]{FFFFFF}CENTEKA   - Industrial}       & \multicolumn{1}{l|}{\cellcolor[HTML]{FFFFFF}SURYAROSNI   - BM}                  & \multicolumn{1}{l|}{\cellcolor[HTML]{FFFFFF}MAHSEAMLES   - BM}                 & \multicolumn{1}{l|}{\cellcolor[HTML]{FFFFFF}JSWSTEEL.NS   - BM}              & HCLTECH.NS   - Technology \\ \cline{2-9}
\cellcolor[HTML]{FFFFFF}                                                               & \multicolumn{1}{l|}{\cellcolor[HTML]{FFFFFF}CANFINHOME   - Finance}      & \multicolumn{1}{l|}{\cellcolor[HTML]{FFFFFF}EIHOTEL   - CC}            & \multicolumn{1}{l|}{\cellcolor[HTML]{FFFFFF}NESTLEIND.NS   - CD}        & \multicolumn{1}{l|}{\cellcolor[HTML]{FFFFFF}SIEMENS   - Industrial}       & \multicolumn{1}{l|}{\cellcolor[HTML]{FFFFFF}KAKATCEM   - BM}                    & \multicolumn{1}{l|}{\cellcolor[HTML]{FFFFFF}RAMANEWS   - BM}                   & \multicolumn{1}{l|}{\cellcolor[HTML]{FFFFFF}BHEL   - Industrial}             & LUPIN.NS   - HC           \\ \cline{2-9}
\cellcolor[HTML]{FFFFFF}                                                               & \multicolumn{1}{l|}{\cellcolor[HTML]{FFFFFF}HARRMALAYA   - CD}           & \multicolumn{1}{l|}{\cellcolor[HTML]{FFFFFF}BAJFINANCE   - Finance}    & \multicolumn{1}{l|}{\cellcolor[HTML]{FFFFFF}DIVISLAB.NS   - HC}         & \multicolumn{1}{l|}{\cellcolor[HTML]{FFFFFF}KSB   - Industrial}           & \multicolumn{1}{l|}{\cellcolor[HTML]{FFFFFF}PIDILITIND   - BM}                  & \multicolumn{1}{l|}{\cellcolor[HTML]{FFFFFF}RELCAPITAL   - Finance}            & \multicolumn{1}{l|}{\cellcolor[HTML]{FFFFFF}DHAMPURSUG   - CD}               & CADILAHC.NS   - HC        \\ \cline{2-9}
\cellcolor[HTML]{FFFFFF}                                                               & \multicolumn{1}{l|}{\cellcolor[HTML]{FFFFFF}RAYMOND   - CC}              & \multicolumn{1}{l|}{\cellcolor[HTML]{FFFFFF}CANFINHOME   - Finance}    & \multicolumn{1}{l|}{\cellcolor[HTML]{FFFFFF}TCS   - Technology}         & \multicolumn{1}{l|}{\cellcolor[HTML]{FFFFFF}TFCILTD   - Finance}          & \multicolumn{1}{l|}{\cellcolor[HTML]{FFFFFF}MARICO.NS   - CD}                   & \multicolumn{1}{l|}{\cellcolor[HTML]{FFFFFF}GIPCL   - Utilities}               & \multicolumn{1}{l|}{\cellcolor[HTML]{FFFFFF}BOMDYING   - CC}                 & ESCORTS   - Industrial    \\ \cline{2-9}
\cellcolor[HTML]{FFFFFF}                                                               & \multicolumn{1}{l|}{\cellcolor[HTML]{FFFFFF}SWARAJENG   - Auto}          & \multicolumn{1}{l|}{\cellcolor[HTML]{FFFFFF}BLUEDART   - Industrial}   & \multicolumn{1}{l|}{\cellcolor[HTML]{FFFFFF}BRITANNIA   - CD}           & \multicolumn{1}{l|}{\cellcolor[HTML]{FFFFFF}HAVELLS.NS   - Industrial}    & \multicolumn{1}{l|}{\cellcolor[HTML]{FFFFFF}COLPAL   - CD}                      & \multicolumn{1}{l|}{\cellcolor[HTML]{FFFFFF}ESABINDIA   - Industrial}          & \multicolumn{1}{l|}{\cellcolor[HTML]{FFFFFF}TATAPOWER   - Utilities}         & AUROPHARMA   - HC         \\ \cline{2-9}
\cellcolor[HTML]{FFFFFF}                                                               & \multicolumn{1}{l|}{\cellcolor[HTML]{FFFFFF}INDIACEM   - BM}             & \multicolumn{1}{l|}{\cellcolor[HTML]{FFFFFF}BAJAJFINSV.NS   - Finance} & \multicolumn{1}{l|}{\cellcolor[HTML]{FFFFFF}MPHASIS.NS   - Technology}  & \multicolumn{1}{l|}{\cellcolor[HTML]{FFFFFF}INDHOTEL   - CC}              & \multicolumn{1}{l|}{\cellcolor[HTML]{FFFFFF}RELCAPITAL   - Finance}             & \multicolumn{1}{l|}{\cellcolor[HTML]{FFFFFF}TFCILTD   - Financial}             & \multicolumn{1}{l|}{\cellcolor[HTML]{FFFFFF}SAIL   - BM}                     & HIMATSEIDE   - CC         \\ \cline{2-9}
\cellcolor[HTML]{FFFFFF}                                                               & \multicolumn{1}{l|}{\cellcolor[HTML]{FFFFFF}GODFRYPHLP   - CD}           & \multicolumn{1}{l|}{\cellcolor[HTML]{FFFFFF}ULTRACEMCO.NS}             & \multicolumn{1}{l|}{\cellcolor[HTML]{FFFFFF}WIPRO   - Technology}       & \multicolumn{1}{l|}{\cellcolor[HTML]{FFFFFF}ULTRACEMCO.NS   - BM}         & \multicolumn{1}{l|}{\cellcolor[HTML]{FFFFFF}BALRAMCHIN   - CD}                  & \multicolumn{1}{l|}{\cellcolor[HTML]{FFFFFF}MIRCELECTR   - CC}                 & \multicolumn{1}{l|}{\cellcolor[HTML]{FFFFFF}GIPCL   - Utilities}             & TATAELXSI   - Technology  \\ \cline{2-9}
\cellcolor[HTML]{FFFFFF}                                                               & \multicolumn{1}{l|}{\cellcolor[HTML]{FFFFFF}BHARATFORG   - Auto}         & \multicolumn{1}{l|}{\cellcolor[HTML]{FFFFFF}UNITECH   - Technology}    & \multicolumn{1}{l|}{\cellcolor[HTML]{FFFFFF}DABUR   - CD}               & \multicolumn{1}{l|}{\cellcolor[HTML]{FFFFFF}CHOLAFIN   - Finance}         & \multicolumn{1}{l|}{\cellcolor[HTML]{FFFFFF}CIPLA   - HC}                       & \multicolumn{1}{l|}{\cellcolor[HTML]{FFFFFF}CANFINHOME   - Finance}            & \multicolumn{1}{l|}{\cellcolor[HTML]{FFFFFF}ADANIENT-   Energy}              & COLPAL   - CD             \\ \cline{2-9}
\multirow{-16}{*}{\rotatebox{90}{\cellcolor[HTML]{FFFFFF}Considering Atmost top 15   positive values}} & \multicolumn{1}{l|}{\cellcolor[HTML]{FFFFFF}TATAMOTORS.NS   - Auto}      & \multicolumn{1}{l|}{\cellcolor[HTML]{FFFFFF}SHREECEM   - BM}           & \multicolumn{1}{l|}{\cellcolor[HTML]{FFFFFF}APOLLOHOSP   - HC}          & \multicolumn{1}{l|}{\cellcolor[HTML]{FFFFFF}ITI   - Technology}           & \multicolumn{1}{l|}{\cellcolor[HTML]{FFFFFF}CADILAHC.NS   - HC}                 & \multicolumn{1}{l|}{\cellcolor[HTML]{FFFFFF}ROLTA   - Technology}              & \multicolumn{1}{l|}{\cellcolor[HTML]{FFFFFF}ZEEL   - CD}                     & RELIANCE   - Financial    \\ \hline
\end{tabular}}
\end{table}

\begin{table}[h]
\caption{Negative influential nodes for different specific target sectors for negative network in post-COVID time, choosing $\beta_0 =0.1, \gamma_0= 0.7$}
\label{table 4}
\resizebox{\columnwidth}{!}{
\begin{tabular}{|c|llllllll|}
\hline
\multicolumn{1}{|l|}{}                                      & \multicolumn{8}{c|}{\textbf{Influential nodes for Different specific target sectors for negative network for post COVID time}}                                                                                                                                                                                                                                                              \\[1ex] \hline
\multirow{16}{*}{\rotatebox{90}{Considering Atmost top 15 positive values}} & \multicolumn{1}{l|}{\textbf{Industrial (Ind)}} & \multicolumn{1}{l|}{\textbf{Financial}}    & \multicolumn{1}{l|}{\textbf{Healthcare (HC)}} & \multicolumn{1}{l|}{\textbf{Automobile  (Auto)}} & \multicolumn{1}{l|}{\textbf{Consumer Defensive (CD)}} & \multicolumn{1}{l|}{\textbf{Consumer Cyclical  (CC)}} & \multicolumn{1}{l|}{\textbf{Basic Materials (BM)}} & \textbf{Technology}   \\ \cline{2-9}
                                                            & \multicolumn{1}{l|}{SHYAMTEL - Technology}     & \multicolumn{1}{l|}{SHYAMTEL - Technology} & \multicolumn{1}{l|}{SHYAMTEL - Technology}    & \multicolumn{1}{l|}{SHYAMTEL - Technology}       & \multicolumn{1}{l|}{SHYAMTEL - Technology}            & \multicolumn{1}{l|}{ESCORTS - Industrial}             & \multicolumn{1}{l|}{SHYAMTEL - Technology}         & SHYAMTEL - Technology \\ \cline{2-9}
                                                            & \multicolumn{1}{l|}{SIYSIL - CC}               & \multicolumn{1}{l|}{ICIL - CC}             & \multicolumn{1}{l|}{HGS - Industrial}         & \multicolumn{1}{l|}{GNFC - BM}                   & \multicolumn{1}{l|}{BAJAJ-AUTO -Auto}                 & \multicolumn{1}{l|}{INFY - Technology}                & \multicolumn{1}{l|}{ESCORTS - Industrial}          & BAJAJ-AUTO -Auto      \\ \cline{2-9}
                                                            & \multicolumn{1}{l|}{ICIL - CC}                 & \multicolumn{1}{l|}{INFY - Technology}     & \multicolumn{1}{l|}{GNFC - BM}                & \multicolumn{1}{l|}{ASIANHOTNR - CC}             & \multicolumn{1}{l|}{}                                 & \multicolumn{1}{l|}{ICIL - CC}                        & \multicolumn{1}{l|}{ICIL - CC}                     & CRISIL - Finance      \\ \cline{2-9}
                                                            & \multicolumn{1}{l|}{HGS - Industrial}          & \multicolumn{1}{l|}{GNFC - BM}             & \multicolumn{1}{l|}{SIYSIL - CC}              & \multicolumn{1}{l|}{TATAELXSI - Technology}      & \multicolumn{1}{l|}{}                                 & \multicolumn{1}{l|}{ASIANHOTNR - CC}                  & \multicolumn{1}{l|}{INFY - Technology}             & LUPIN.NS - HC         \\ \cline{2-9}
                                                            & \multicolumn{1}{l|}{APOLLOHOSP - HC}           & \multicolumn{1}{l|}{HGS - Industrial}      & \multicolumn{1}{l|}{INFY - Technology}        & \multicolumn{1}{l|}{ELGIEQUIP - Industrial}      & \multicolumn{1}{l|}{}                                 & \multicolumn{1}{l|}{CHOLAFIN - Finance}               & \multicolumn{1}{l|}{BAJAJ-AUTO - Auto}                    & FINPIPE - Industrial  \\ \cline{2-9}
                                                            & \multicolumn{1}{l|}{ASIANHOTNR - CC}           & \multicolumn{1}{l|}{M\&M - Auto}           & \multicolumn{1}{l|}{BERGERPAINT - BM}         & \multicolumn{1}{l|}{BAJAJ-AUTO -Auto}            & \multicolumn{1}{l|}{}                                 & \multicolumn{1}{l|}{BAJAJ-AUTO -Auto}                 & \multicolumn{1}{l|}{HGS - Industrial}              & BATAINDIA - CD        \\ \cline{2-9}
                                                            & \multicolumn{1}{l|}{INFY - Technology}         & \multicolumn{1}{l|}{}                      & \multicolumn{1}{l|}{ESCORTS - Industrial}     & \multicolumn{1}{l|}{FOSECOIND - BM}              & \multicolumn{1}{l|}{}                                 & \multicolumn{1}{l|}{M\&M - Auto}                      & \multicolumn{1}{l|}{APOLLOHOSP - HC}               & AUROPHARMA - HC       \\ \cline{2-9}
                                                            & \multicolumn{1}{l|}{BAJAJ-AUTO -Auto}          & \multicolumn{1}{l|}{}                      & \multicolumn{1}{l|}{}                         & \multicolumn{1}{l|}{ICIL - CC}                   & \multicolumn{1}{l|}{}                                 & \multicolumn{1}{l|}{APOLLOHOSP - HC}                  & \multicolumn{1}{l|}{TATAELXSI - Technology}        & MRF - Auto            \\ \cline{2-9}
                                                            & \multicolumn{1}{l|}{NAGAFERT - BM}             & \multicolumn{1}{l|}{}                      & \multicolumn{1}{l|}{}                         & \multicolumn{1}{l|}{TNPETRO - BM}                & \multicolumn{1}{l|}{}                                 & \multicolumn{1}{l|}{BERGERPAINT - BM}                 & \multicolumn{1}{l|}{PFIZER - HC}                   & APOLLOHOSP - HC       \\ \cline{2-9}
                                                            & \multicolumn{1}{l|}{GESHIP - Industrial}       & \multicolumn{1}{l|}{}                      & \multicolumn{1}{l|}{}                         & \multicolumn{1}{l|}{IFCI - Financial}            & \multicolumn{1}{l|}{}                                 & \multicolumn{1}{l|}{PFIZER - HC}                      & \multicolumn{1}{l|}{M\&M - Auto}                   & GNFC - BM             \\ \cline{2-9}
                                                            & \multicolumn{1}{l|}{IPCALAB - HC}              & \multicolumn{1}{l|}{}                      & \multicolumn{1}{l|}{}                         & \multicolumn{1}{l|}{CENTURYTEX - BM}             & \multicolumn{1}{l|}{}                                 & \multicolumn{1}{l|}{MAHSCOOTER - Auto}                & \multicolumn{1}{l|}{EICHERMOT - Auto}              & MAHSCOOTER - Auto     \\ \cline{2-9}
                                                            & \multicolumn{1}{l|}{MAHSEAMLES - BM}           & \multicolumn{1}{l|}{}                      & \multicolumn{1}{l|}{}                         & \multicolumn{1}{l|}{HCLTECH.NS - Technology}     & \multicolumn{1}{l|}{}                                 & \multicolumn{1}{l|}{SUNDRMFAST - Auto}                & \multicolumn{1}{l|}{SBIN - Finance}                & PFIZER - HC           \\ \cline{2-9}
                                                            & \multicolumn{1}{l|}{BASF - BM}                 & \multicolumn{1}{l|}{}                      & \multicolumn{1}{l|}{}                         & \multicolumn{1}{l|}{SHANTIGEAR - Auto}           & \multicolumn{1}{l|}{}                                 & \multicolumn{1}{l|}{LICHSGFIN - Finance}              & \multicolumn{1}{l|}{MARUTI.NS - Auto}              & FMGOETZE - Auto       \\ \cline{2-9}
                                                            & \multicolumn{1}{l|}{MAHSCOOTER - Auto}         & \multicolumn{1}{l|}{}                      & \multicolumn{1}{l|}{}                         & \multicolumn{1}{l|}{LUMAXIND - Auto}             & \multicolumn{1}{l|}{}                                 & \multicolumn{1}{l|}{NAHARSPING - CC}                  & \multicolumn{1}{l|}{MAHSCOOTER - Auto}             & CHOLAFIN - Finance    \\ \cline{2-9}
                                                            & \multicolumn{1}{l|}{HIMATSEIDE - CC}           & \multicolumn{1}{l|}{}                      & \multicolumn{1}{l|}{}                         & \multicolumn{1}{l|}{CARBORUNIV - Industrial}     & \multicolumn{1}{l|}{}                                 & \multicolumn{1}{l|}{EIHOTEL - CC}                     & \multicolumn{1}{l|}{ASIANHOTNR - CC}               & NESTLEIND.NS - CD     \\ \hline
\end{tabular}}
\end{table}
\end{landscape}

\begin{figure}[H]
\centering
\begin{tabular}{cccc}
\includegraphics[width=0.25\textwidth]{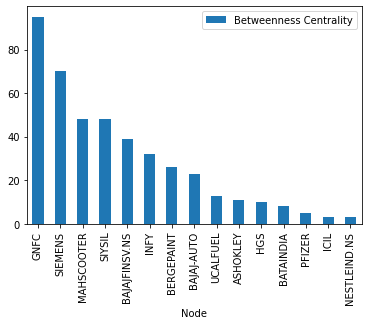} &
\includegraphics[width=0.25\textwidth]{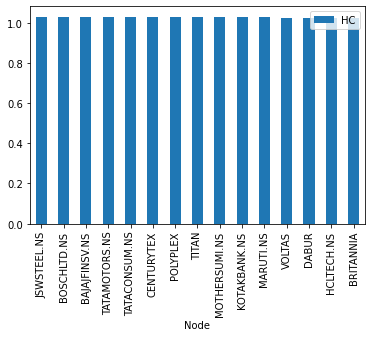} &
\includegraphics[width=0.25\textwidth]{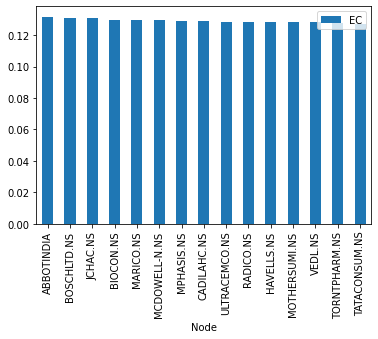} \\
\textbf{(a)}  & \textbf{(b)} & \textbf{(c)}  \\[6pt]
\end{tabular}
\begin{tabular}{cccc}
\includegraphics[width=0.25\textwidth]{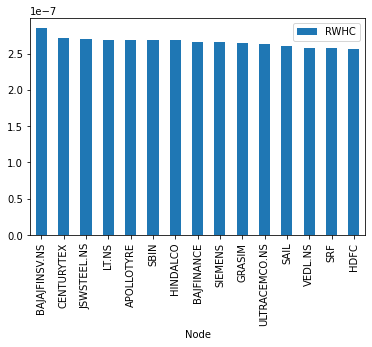} &
\includegraphics[width=0.25\textwidth]{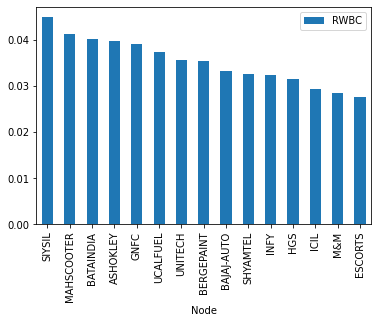} \\
\textbf{(d)}  & \textbf{(e)}     \\[6pt]
\end{tabular}
\caption{Top 15 nodes using \textbf{(a)} Betweenness centrality measures
\textbf{(b)} Harmonic centrality measures
\textbf{(c)} Eigen centrality measures
\textbf{(d)} Random walk harmonic centrality measures
\textbf{(e)} Random walk betweenness centrality, for positive network during in-COVID time period.}
\label{fig: Global centrality measures pos incovid}

\centering
\begin{tabular}{cccc}
\includegraphics[width=0.25\textwidth]{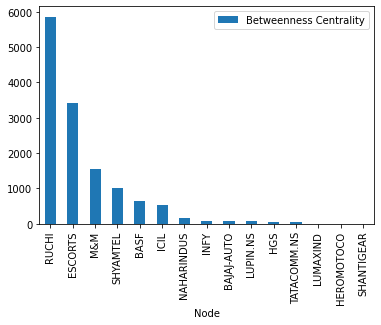} &
\includegraphics[width=0.25\textwidth]{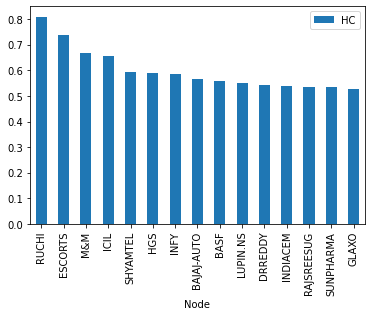} &
\includegraphics[width=0.25\textwidth]{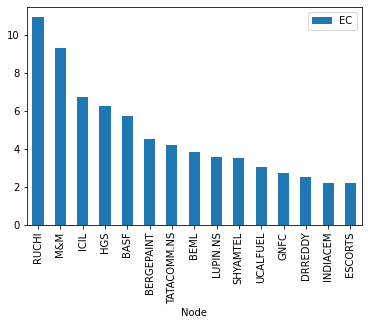} \\
\textbf{(a)}  & \textbf{(b)} & \textbf{(c)}  \\[6pt]
\end{tabular}
\begin{tabular}{cccc}
\includegraphics[width=0.25\textwidth]{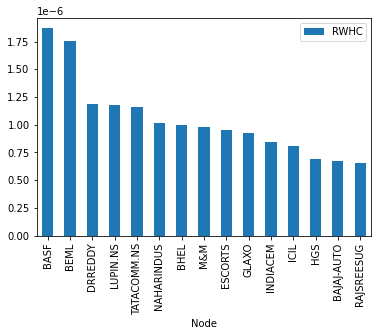} &
\includegraphics[width=0.25\textwidth]{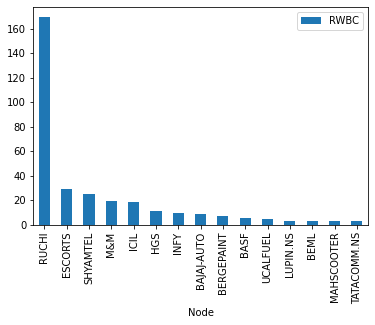} \\
\textbf{(d)} & \textbf{(e)}   \\[6pt]
\end{tabular}
\caption{ Top 15 nodes using \textbf{(a)} Betweenness centrality measures
\textbf{(b)} Harmonic centrality measures
\textbf{(c)} Eigen centrality measures
\textbf{(d)} Random walk harmonic centrality measures
\textbf{(e)} Random walk betweenness centrality, for negative network during in-COVID time period.}
\label{fig: Global centrality measures neg incovid}
\end{figure}

\begin{figure}[H]
\centering
\begin{tabular}{cccc}
\includegraphics[width=0.25\textwidth]{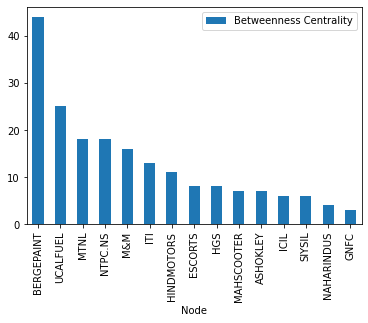} &
\includegraphics[width=0.25\textwidth]{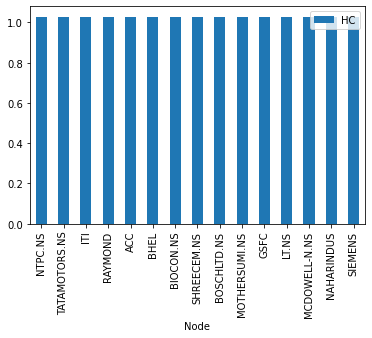} &
\includegraphics[width=0.25\textwidth]{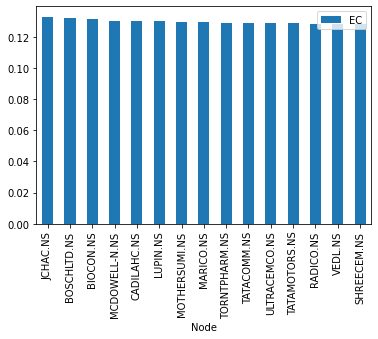} \\
\textbf{(a)}  & \textbf{(b)} & \textbf{(c)}  \\[6pt]
\end{tabular}
\begin{tabular}{cccc}
\includegraphics[width=0.25\textwidth]{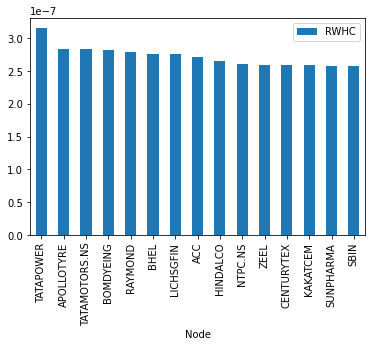} &
\includegraphics[width=0.25\textwidth]{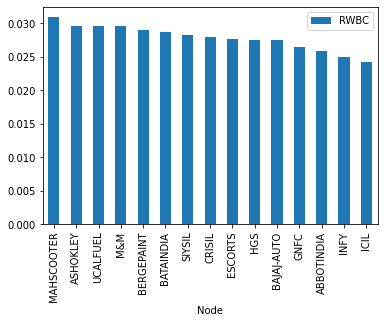} \\
\textbf{(d)}  & \textbf{(e)}     \\[6pt]
\end{tabular}
\caption{Top 15 nodes using \textbf{(a)} Betweenness centrality measures
\textbf{(b)} Harmonic centrality measures
\textbf{(c)} Eigen centrality measures
\textbf{(d)} Random walk harmonic centrality measures
\textbf{(e)} Random walk betweenness centrality, for positive network in post-COVID time period.}
\label{fig: Global centrality measures pos postcovid}

\centering
\begin{tabular}{cccc}
\includegraphics[width=0.25\textwidth]{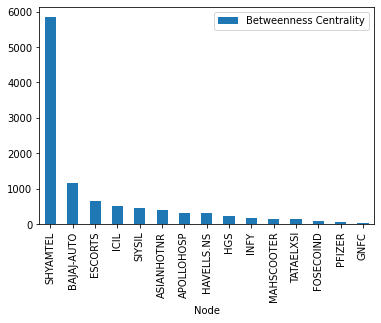} &
\includegraphics[width=0.25\textwidth]{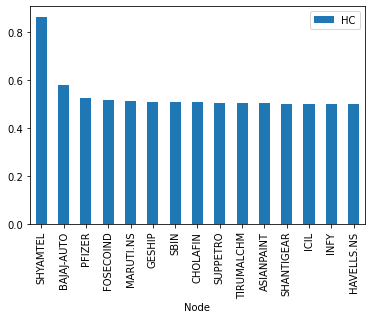} &
\includegraphics[width=0.25\textwidth]{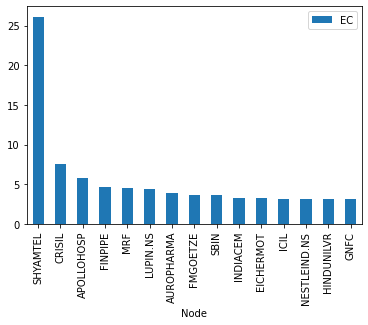} \\
\textbf{(a)}  & \textbf{(b)} & \textbf{(c)}  \\[6pt]
\end{tabular}
\begin{tabular}{cccc}
\includegraphics[width=0.25\textwidth]{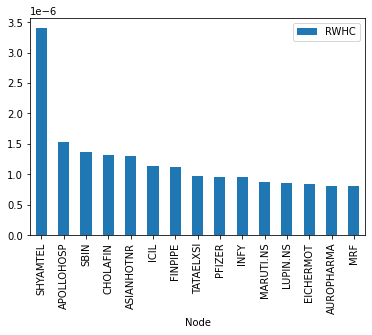} &
\includegraphics[width=0.25\textwidth]{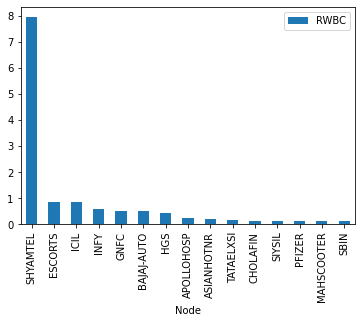} \\
\textbf{(d)} & \textbf{(e)}   \\[6pt]
\end{tabular}
\caption{Top 15 nodes using \textbf{(a)} Betweenness centrality measures
\textbf{(b)} Harmonic centrality measures
\textbf{(c)} Eigen centrality measures
\textbf{(d)} Random walk harmonic centrality measures
\textbf{(e)} Random walk betweenness centrality, for negative network in post-COVID time period.}
\label{fig: Global centrality measures neg postcovid}
\end{figure}

\section{Summary and Conclusion:}
\label{sec: conclude}

The LIEST methodology (See Table \ref{table 1}- \ref{table 4}) has been employed in this study for the analysis of influential effects separately for positive and negative cases during the in and post-COVID periods, where the time series data used are the daily stock prices of different companies trading stocks in the National Stock Exchange of India. The present work gives a detailed account of the positive and negative influences which is extremely relevant and contextual in real-life situations.The active participation of Health Care and Consumer Defensive sectors along with Financial, Industrial, Technology and Automobile sectors with reduced participation of Consumer Cyclical and Basic Materials sectors have been observed with positive effect for different sectors during the in-COVID period. On the other hand, during the recovery phase of the post-COVID period the Financial sectors along with other sectors have shown their positive impact on others. Increased active participation of Healthcare, Consumer Defensive and Industrial sectors also created a negative impact for Automobile and Technology sectors during the in-COVID time. During the recovery span, the Automobile and Technology sectors largely created negative effect on most of the sectors. \\

Different global centrality measures (see Fig. \ref{fig: Global centrality measures pos incovid} - \ref{fig: Global centrality measures neg postcovid} ) such as Betweenness Centrality (BC), Harmonic Centrality (HC), Eigen Centrality, Random walk Betweenness Centrality (RWBC)\cite{RWBC}, Random walk Harmonic Centrality (RWHC) have been used to study the overall impact of the companies on the market.Though the impact of all sectors can be observed through the results of global centrality measures, we mention here a few important sectors belonging  both to positive and negative networks during the in-COVID and post-COVID periods. From the network analysis using global centrality measures (see Fig. \ref{fig: Global centrality measures pos incovid} - \ref{fig: Global centrality measures neg postcovid} ), the influences of different sectors can be visualized but it is not possible to distinguish between the positive and the negative influences. In the present analysis, attempts have been made to overcome the limitations due to the global centrality measures. \\

It has been observed that aggressiveness, conservatism, overconfidence and similar sentiments, which are the consequences of mass psychology, also influence the stock market to a large extent \cite{baker2007psychology, de1990psychology}. These psychological consequences along with the index of the stock market directly or indirectly affects the similarity measures between the stocks which has also been reflected in the present study. \\

As the order and complexity of a network increases, local influences (in the $k-neighbourhood$ of a vertex or those received from a specific subset of vertices in the network) may be important to study in order to understand the causal flows acting on a particular vertex. It is thus possible using LIEST to identify which pair of companies can potentially have a causality relation or not. 
The use of pattern causality (or some other method) may be suggested to find causality only between those pairs of companies for which the causality relationship might have been established using LIEST. The idea of pattern causality and related techniques will be explored by the present authors in a subsequent publication.

\section*{Credit Author Statement}
Authors AS and PKP conceptualized the idea upon discussion which was emphasized by authors SU and IM. Author AS designed the methodology. Author AS collected, systemically organised and analysed the data, visualised and interpreted the results with supervision from authors SU, IM and PKP. Author AS wrote the draft of the manuscript which was revised and edited by authors SU, IM and PKP. 

\section*{Declaration of competing interest}
The authors declare that they have no known competing financial interests or personal relationships that could have appeared to influence the work reported in this paper.

\section*{Acknowledgement} 

Author SU would like to acknowledge the financial support received under the project ``Quantum information technologies with photonic devices (DST)'' (Ref. No.: IISER-K/DoRD/R\&P/2021-22/425) funded by QuEST, DST. Authors AS and IM would like to acknowledge the infrastructural and computational support provided by MAKAUT, WB during preparation of the manuscript.

\begin{table}[]
\caption{The list of 175 stocks in NSE analyzed in this paper.}
\label{9}
\resizebox{\textwidth}{!}{
\begin{tabular}{|r|l|l|l|r|l|l|l|r|l|l|}
\hline
\multicolumn{1}{|l|}{\textbf{i}} & \textbf{Company} & \textbf{Sector} &  & \multicolumn{1}{l|}{\textbf{i}} & \textbf{Company} & \textbf{Sector}   &  & \multicolumn{1}{l|}{\textbf{i}} & \textbf{Company} & \textbf{Sector}    \\ \hline
1                                & ESCORTS          & Industrial      &  & 61                              & BERGEPAINT       & Basic Materials   &  & 121                             & BOSCHLTD.NS      & Automobile         \\ \hline
2                                & ASHOKLEY         & Industrial      &  & 62                              & GNFC             & Basic Materials   &  & 122                             & IDBI             & Financial          \\ \hline
3                                & BEML             & Industrial      &  & 63                              & DEEPAKFERT       & Basic Materials   &  & 123                             & HDFCBANK         & Financial          \\ \hline
4                                & RIIL             & Industrial      &  & 64                              & GSFC             & Basic Materials   &  & 124                             & SBIN             & Financial          \\ \hline
5                                & FINPIPE          & Industrial      &  & 65                              & SRF              & Basic Materials   &  & 125                             & KARURVYSYA       & Financial          \\ \hline
6                                & VESUVIUS         & Industrial      &  & 66                              & GRASIM           & Basic Materials   &  & 126                             & IFCI             & Financial          \\ \hline
7                                & INGERRAND        & Industrial      &  & 67                              & INDIACEM         & Basic Materials   &  & 126                             & RELCAPITAL       & Financial          \\ \hline
8                                & ELGIEQUIP        & Industrial      &  & 68                              & KESORAMIND       & Basic Materials   &  & 128                             & CHOLAFIN         & Financial          \\ \hline
9                                & KSB              & Industrial      &  & 69                              & RAMANEWS         & Basic Materials   &  & 129                             & BAJFINANCE       & Financial          \\ \hline
10                               & VOLTAS           & Industrial      &  & 70                              & POLYPLEX         & Basic Materials   &  & 130                             & HDFC             & Financial          \\ \hline
11                               & KEC              & Industrial      &  & 71                              & SPIC             & Basic Materials   &  & 131                             & LICHSGFIN        & Financial          \\ \hline
12                               & BHEL             & Industrial      &  & 72                              & NAGAFERT         & Basic Materials   &  & 132                             & CANFINHOME       & Financial          \\ \hline
13                               & THERMAX          & Industrial      &  & 73                              & ACC              & Basic Materials   &  & 133                             & GICHSGFIN        & Financial          \\ \hline
14                               & HEG              & Industrial      &  & 74                              & CENTURYTEX       & Basic Materials   &  & 134                             & TFCILTD          & Financial          \\ \hline
15                               & ESABINDIA        & Industrial      &  & 75                              & JSWSTEEL.NS      & Basic Materials   &  & 135                             & CRISIL           & Financial          \\ \hline
16                               & CENTENKA         & Industrial      &  & 76                              & SHREECEM.NS      & Basic Materials   &  & 136                             & BAJAJFINSV.NS    & Financial          \\ \hline
17                               & HGS              & Industrial      &  & 77                              & ULTRACEMCO.NS    & Basic Materials   &  & 137                             & KOTAKBANK.NS     & Financial          \\ \hline
18                               & CARBORUNIV       & Industrial      &  & 78                              & VEDL.NS          & Basic Materials   &  & 138                             & ITC              & Consumer Defensive \\ \hline
19                               & SUPREMEIND       & Industrial      &  & 79                              & HINDPETRO        & Energy            &  & 139                             & VSTIND           & Consumer Defensive \\ \hline
20                               & GESHIP           & Industrial      &  & 80                              & ONGC             & Energy            &  & 140                             & GODFRYPHLP       & Consumer Defensive \\ \hline
21                               & FINCABLES        & Industrial      &  & 81                              & RELIANCE         & Energy            &  & 141                             & HARRMALAYA       & Consumer Defensive \\ \hline
22                               & BLUEDART         & Industrial      &  & 82                              & BPCL             & Energy            &  & 142                             & BALRAMCHIN       & Consumer Defensive \\ \hline
23                               & ABB              & Industrial      &  & 83                              & ADANIENT         & Energy            &  & 143                             & RAJSREESUG       & Consumer Defensive \\ \hline
24                               & SIEMENS          & Industrial      &  & 84                              & NTPC.NS          & Energy            &  & 144                             & SAKHTISUG        & Consumer Defensive \\ \hline
25                               & LT.NS            & Industrial      &  & 85                              & MIRCELECTR       & Consumer Cyclical &  & 145                             & DHAMPURSUG       & Consumer Defensive \\ \hline
26                               & HAVELLS.NS       & Industrial      &  & 86                              & BATAINDIA        & Consumer Cyclical &  & 146                             & BRITANNIA        & Consumer Defensive \\ \hline
27                               & GLAXO            & Healthcare      &  & 87                              & ICIL             & Consumer Cyclical &  & 147                             & RUCHI            & Consumer Defensive \\ \hline
28                               & DRREDDY          & Healthcare      &  & 88                              & ARVIND           & Consumer Cyclical &  & 148                             & DABUR            & Consumer Defensive \\ \hline
29                               & CIPLA            & Healthcare      &  & 89                              & RAYMOND          & Consumer Cyclical &  & 149                             & COLPAL           & Consumer Defensive \\ \hline
30                               & SUNPHARMA        & Healthcare      &  & 90                              & HIMATSEIDE       & Consumer Cyclical &  & 150                             & HINDUNILVR       & Consumer Defensive \\ \hline
31                               & IPCALAB          & Healthcare      &  & 91                              & BOMDYEING        & Consumer Cyclical &  & 151                             & EIDPARRY         & Consumer Defensive \\ \hline
32                               & PFIZER           & Healthcare      &  & 92                              & NAHARSPING       & Consumer Cyclical &  & 152                             & ZEEL             & Consumer Defensive \\ \hline
33                               & AUROPHARMA       & Healthcare      &  & 93                              & MARALOVER        & Consumer Cyclical &  & 153                             & NESTLEIND.NS     & Consumer Defensive \\ \hline
34                               & NATCOPHARM       & Healthcare      &  & 94                              & SIYSIL           & Consumer Cyclical &  & 154                             & TATACONSUM.NS    & Consumer Defensive \\ \hline
35                               & APOLLOHOSP       & Healthcare      &  & 95                              & INDHOTEL         & Consumer Cyclical &  & 155                             & RADICO.NS        & Consumer Defensive \\ \hline
36                               & DIVISLAB.NS      & Healthcare      &  & 96                              & EIHOTEL          & Consumer Cyclical &  & 156                             & MCDOWELL-N.NS    & Consumer Defensive \\ \hline
37                               & TORNTPHARM.NS    & Healthcare      &  & 97                              & ASIANHOTNR       & Consumer Cyclical &  & 157                             & MARICO.NS        & Consumer Defensive \\ \hline
38                               & CADILAHC.NS      & Healthcare      &  & 98                              & COSMOFILMS       & Consumer Cyclical &  & 158                             & TATAELXSI        & Technology         \\ \hline
39                               & BIOCON.NS        & Healthcare      &  & 99                              & THOMASCOOK       & Consumer Cyclical &  & 159                             & ROLTA            & Technology         \\ \hline
40                               & ABBOTINDIA       & Healthcare      &  & 100                             & TITAN            & Consumer Cyclical &  & 160                             & INFY             & Technology         \\ \hline
41                               & LUPIN.NS         & Healthcare      &  & 101                             & NAHARINDUS       & Consumer Cyclical &  & 161                             & MASTEK           & Technology         \\ \hline
42                               & TNPETRO          & Basic Materials &  & 102                             & JCHAC.NS         & Consumer Cyclical &  & 162                             & WIPRO            & Technology         \\ \hline
43                               & SUPPETRO         & Basic Materials &  & 103                             & LUMAXIND         & Automobile        &  & 163                             & SHYAMTEL         & Technology         \\ \hline
44                               & DCW              & Basic Materials &  & 104                             & HEROMOTOCO       & Automobile        &  & 164                             & BIRLACABLE       & Technology         \\ \hline
45                               & NOCIL            & Basic Materials &  & 105                             & SHANTIGEAR       & Automobile        &  & 165                             & TCS              & Technology         \\ \hline
46                               & TIRUMALCHM       & Basic Materials &  & 106                             & MAHSCOOTER       & Automobile        &  & 166                             & ITI              & Technology         \\ \hline
47                               & TATACHEM         & Basic Materials &  & 107                             & BAJAJ-AUTO       & Automobile        &  & 167                             & HCLTECH.NS       & Technology         \\ \hline
48                               & GHCL             & Basic Materials &  & 108                             & EICHERMOT        & Automobile        &  & 168                             & MPHASIS.NS       & Technology         \\ \hline
49                               & GUJALKALI        & Basic Materials &  & 109                             & HINDMOTORS       & Automobile        &  & 169                             & GIPCL            & Utilities          \\ \hline
50                               & PIDILITIND       & Basic Materials &  & 110                             & SWARAJENG        & Automobile        &  & 170                             & CESC             & Utilities          \\ \hline
51                               & FOSECOIND        & Basic Materials &  & 111                             & APOLLOTYRE       & Automobile        &  & 171                             & TATAPOWER        & Utilities          \\ \hline
52                               & BASF             & Basic Materials &  & 112                             & FMGOETZE         & Automobile        &  & 172                             & UNITECH          & Real Estate        \\ \hline
53                               & HINDALCO         & Basic Materials &  & 113                             & MRF              & Automobile        &  & 173                             & MTNL             & Telecom            \\ \hline
54                               & SAIL             & Basic Materials &  & 114                             & UCALFUEL         & Automobile        &  & 174                             & BHARTIARTL.NS    & Telecom            \\ \hline
55                               & TATAMETALI       & Basic Materials &  & 115                             & BHARATFORG       & Automobile        &  & 175                             & TATACOMM.NS      & Telecom            \\ \hline
56                               & MAHSEAMLES       & Basic Materials &  & 116                             & M\&M             & Automobile        &  & \multicolumn{1}{l|}{}           &                  &                    \\ \hline
57                               & SURYAROSNI       & Basic Materials &  & 117                             & SUNDRMFAST       & Automobile        &  & \multicolumn{1}{l|}{}           &                  &                    \\ \hline
58                               & TNPL             & Basic Materials &  & 118                             & MARUTI.NS        & Automobile        &  & \multicolumn{1}{l|}{}           &                  &                    \\ \hline
59                               & KAKATCEM         & Basic Materials &  & 119                             & TATAMOTORS.NS    & Automobile        &  & \multicolumn{1}{l|}{}           &                  &                    \\ \hline
60                               & ASIANPAINT       & Basic Materials &  & 120                             & MOTHERSUMI.NS    & Automobile        &  & \multicolumn{1}{l|}{}           &                  &                    \\ \hline
\end{tabular}}
\end{table}

\pagebreak
\bibliographystyle{plain}
\bibliography{liest.bib}

\end{document}